\documentclass[aps,prd,twocolumn,showpacs,amsmath,floatfix]{revtex4}

\usepackage{amsmath}
\usepackage{amssymb}
\usepackage{bm}
\usepackage{graphicx}
\usepackage{rotating}
\usepackage{hyperref}

\newcommand{\be}{\begin{equation}}
\newcommand{\ee}{\end{equation}}
\newcommand{\beq}{\begin{equation}}
\newcommand{\eeq}{\end{equation}}
\newcommand{\bea}{\begin{eqnarray}}
\newcommand{\eea}{\end{eqnarray}}
\newcommand{\bdm}{\begin{displaymath}}
\newcommand{\edm}{\end{displaymath}}

\begin{document}
\title{LISA detections of massive black hole inspirals: parameter extraction errors due to inaccurate template waveforms}  
\author{Curt Cutler}
\affiliation{Jet Propulsion Laboratory, California Institute of Technology, Pasadena, CA 91109}
\author{Michele Vallisneri}
\affiliation{Jet Propulsion Laboratory, California Institute of Technology, Pasadena, CA 91109}
\date{\today}

\begin{abstract}
The planned Laser Interferometer Space Antenna (LISA) is expected to detect
the inspiral and merger of massive black hole binaries (MBHBs) at $z \lesssim 5$ 
with signal-to-noise ratios (SNRs) of hundreds to thousands.  Because of these high SNRs, and because these SNRs accrete over periods of weeks to months, it should be possible to extract the physical parameters of these systems with high accuracy; for instance, for a $\sim 10^6 M_{\odot}$ MBHBs at $z=1$ it should be possible to determine the two masses to $\sim 0.1\%$ and the sky location to $\sim 1^{\circ}$.  However, those are just the errors due to noise: there will be 
additional ``theoretical'' errors due to inaccuracies in our best model waveforms, which are still only approximate. The goal of this paper is to estimate the typical magnitude of these theoretical errors.  We develop mathematical tools for this purpose, and apply them to a somewhat simplified version of the MBHB problem, in which we consider just the inspiral part of the waveform and neglect spin-induced precession, eccentricity, and PN amplitude corrections.  For this simplified version, we estimate that theoretical uncertainties in sky position will typically be $\sim 1^{\circ}$, i.e., comparable to the statistical uncertainty. 
For the mass and spin parameters, our results suggest that while theoretical errors will be rather small absolutely, they
could still dominate over statistical errors (by roughly an order of magnitude) for the strongest sources. 
The tools developed here should be useful for estimating the magnitude of theoretical errors in many other problems in gravitational-wave astronomy.
\end{abstract}
\maketitle

\section{Introduction}
\label{sec:introduction}

The inspirals and mergers of massive ($\sim 10^6 M_{\odot}$) black-hole binaries (MBHBs) 
are likely to be the strongest gravitational-wave (GW) sources detected by LISA, the planned Laser Interferometer  Space Antenna. 
MBHBs at redshifts $z \lesssim 5$ will be detected by LISA with optimal matched-filtering signal-to-noise ratios (SNRs) $\sim 10^2\mbox{--}10^3$.  Because of these high SNRs, and because these sources emit GWs in the LISA band for weeks to months, it should be possible to determine the source parameters to very high accuracy.  For instance, from the inspiral waveform of MBHB systems at $z=1$ it should be possible to extract the two masses to within $\sim 0.1\%$, the spins to within $ \sim 0.1\mbox{--}1\%$, the luminosity distance $D_L$ to $ \sim 0.1\mbox{--}1\% $, and the sky location to within $\sim 0.1\mbox{--}1^{\circ}$ \cite{Lang:2006bz}.
  
More precisely, these are the sizes of the \emph{statistical} errors due to random noise, as calculated using the
Fisher-matrix formalism and (over the years) increasingly faithful models (\emph{templates}) of the gravitational waveforms (see, e.g., Refs.\ \cite{Cutler1998PhRvD..57.7089C,Vecchio2004PhRvD..70d2001V,Sintes2000gr.qc.....5058S,Lang:2006bz}; see also Ref.\ \cite{Vallisneri:2007} about the conditions required for the Fisher-matrix formalism to be applicable). While waveform models have improved, they are still only approximate versions of the true general-relativistic (GR) waveforms, and using them to fit the detector data will incur an additional \emph{theoretical} error in estimating source parameters. 
Note that, to lowest order, while the statistical error due to noise scales roughly as $1/\mathrm{SNR}$, the theoretical error will be
independent of SNR.  For precisely this reason, we might well worry that for the highest SNR sources -- and especially MBHB inspirals -- theoretical errors could dominate the total parameter-estimation error. This basic point has already been
made by Berti \cite{Berti2006CQGra..23S.785B}; however we believe that this paper 
provides the first fairly realistic estimates of the likely magnitude of such 
theoretical errors. 

In recent years numerical relativity (NR) has made great strides toward solving Einstein's equations for the final stages of MBHB inspirals, including the merger and ringdown phases \cite{2005PhRvL..95l1101P,2006PhRvL..96k1101C,2006PhRvL..96k1102B}, and work has begun on the comparison of NR waveforms with the last cycles of post-Newtonian (PN) waveforms \cite{Berti0703053,pan07041964}.
Still, the PN expansion is still the best available method for calculating the vast majority of the $\sim 10^4\mbox{--}10^5$ GW cycles observed by LISA, and it is likely to remain so for many years. In the PN approximation, the GR solution is written as a
``Newtonian'' solution plus corrections given as a power series in $v/c$~\cite{Weinberg}.
Our convention is to refer to corrections of order $(v/c)^{k}$ to either the instantaneous 
orbital elements or their decay rates as $k\mathrm{PN}$ corrections.
Thus we refer to the lowest-order waveform, corresponding to a quasi-Newtonian orbit
that is slowly decaying according to the energy-loss rate given by the quadrupole formula, 
as  the 0PN or Newtonian waveform. 
For circular nonspinning BHs, approximate waveforms have been calculated to 3.5PN order \cite{Blanchet:2001ax}.
For BHs with significant spins (which cause the orbital plane to precess if the spin axes are not exactly aligned
with the binary's orbital angular momentum), the state of the art is somewhat less advanced, as the corresponding
waveforms have been calculated through 2.5PN order \cite{Faye2006PhRvD..74j4033F,Blanchet2006PhRvD..74j4034B}.
Unfortunately there is as yet no general algorithm for extending
PN calculations to arbitrarily high order, and each new order appears to require much additional effort.

We were particularly motivated in researching this paper by the question of how much LISA's angular
resolution is limited by the accuracy of the best available PN templates. This question is important because feasible searches for electromagnetic 
counterparts to gravitationally observed MBHB mergers are likely to require GW-derived error boxes no larger than a 
few square degrees on the sky.  The statistical errors in sky position will generally satisfy this, but what about theoretical errors?
Put another way:  given the great effort and cost of building LISA,  
it seems that a reasonable goal for waveform-modeling theorists
is that parameter-estimation errors due to inaccurate templates should
be no larger than those arising from noise.  
This paper develops tools that will help theorists estimate how close they currently are to this goal,
 where future effort might best be placed, and when they can reasonably stop.   

At this point, the reader might well be wondering: if full waveforms $h_\mathrm{GR}$ from NR are not yet available, by what comparison can we judge the accuracy of high-order PN solutions? Strictly speaking, of course we cannot; nevertheless, there at least two
ways that we can explore this problem.  First, in the limit of very small mass ratios ($M_2/M_1 \ll 1$), the smaller BH can be treated as
a perturbation on the spacetime of the larger BH, and the problem can be solved up to order  $(M_2/M_1)^2$ using BH perturbation theory.
This is straightforward when both BHs are nonspinning, so at least in this case we can compare the high-order PN solution to a solution
of (practically) arbitrary accuracy.

Second, since the PN expansion is known to converge rather slowly for the late stages of inspiral \cite{2001PhRvD..63d4023D,2002PhRvD..66b7502D}, 
we might estimate that the waveform error $h_\mathrm{GR}(t) -  h_{3.5\mathrm{PN}}(t)$  is comparable in size to $h_{3.5\mathrm{PN}}(t) - h_{3.0\mathrm{PN}}(t)$. 
The latter observation leads to a particularly simple method for estimating the magnitude of theoretical errors, which we
implement in this paper.  In effect, we ``replace'' the true waveform $h_\mathrm{GR}(t)$ by $h_{3.5\mathrm{PN}}(t)$ and its
best approximation $h_{3.5\mathrm{PN}}(t)$ by $h_{3.0\mathrm{PN}}(t)$, and directly calculate the resulting errors.   Indeed, essentially this same starting point was taken by  
Canitrot \cite{Canitrot}, who investigated theoretical errors in the context of stellar-mass compact binaries detected by ground-based GW detectors.
Canitrot found, e.g., that fitting 2.0PN templates to 2.5PN waveforms (using the frequency-weighting appropriate for the Virgo detector noise curve) leads to 
theoretical errors in the chirp mass of order $\sim 1\%$ (though small, this is larger than the estimated statistical error), and errors in the symmetric mass ratio of order $\sim 50\%$.

While we believe that this method leads to a useful first estimate of theoretical errors, it ignores data-analysis strategies that might be used to mitigate them.
In this paper, we investigate two such methods of reducing error.  The first involves deweighting the waveform in the last stages of inspiral, where the PN approximation (at any given order) is most uncertain.  The second strategy 
involves using as templates \emph{hybrid} waveforms that include all available \emph{test-mass} PN corrections, which are known to much higher PN order than the corresponding comparable-mass terms.

The plan of this paper is as follows. In Sec.\ \ref{sec:parameterestimation} we give a
rather general geometrical description of 
parameter-estimation errors in GW data analysis, and we develop  
two tools for estimating the sizes of theoretical errors.  The first tool relies on the integration of an ordinary differential equation (ODE) to find the approximate template that best fits a given true GW signal (and vice versa).
The second tool is a one-step substitute for the ODE, which
is generally accurate enough for our purposes and considerably faster to evaluate.
We feel that both of these tools could be profitably applied to 
many similar problems, especially in ground-based
GW astronomy. 
In Sec.\ \ref{sec:sigmodel} we describe the PN waveforms that we adopt as our model waveforms for MBHB inspirals.  Since in this paper we attempt only a first pass at the
full problem, for simplicity a) we employ the restricted PN approximation, which retains
high-order PN corrections to the phase of the waveform's quadrupole component, but neglects both 
higher-multipole components and PN corrections to  the waveform's amplitude;
b) we neglect the spin-induced precession of the orbital plane; and
c) we neglect the (post-inspiral) merger and ringdown phases of the waveform.  
It should be straightforward (if laborious) to restore these neglected pieces in future work.
In Sec.\ \ref{sec:sigmodel} we also summarize our treatment of the LISA response function, which relies on a low-frequency approximation \cite{Cutler1998PhRvD..57.7089C} that is quite adequate for our
present purposes.
Our results are presented in Sec.\ \ref{sec:results}. In Sec.\ \ref{sec:conclusion} we summarize our conclusions and list some directions for future  research.

Throughout this paper, we use geometrical units in which $G=c=1$. Therefore everything can be measured in a fundamental unit of seconds.   For familiarity, we sometimes express 
quantities in terms of yr, Mpc, or $M_\odot$, which are related to our fundamental unit by 1 yr $= 3.1556 \times 10^7$ s, 1 Mpc $= 1.029 \times 10^{14}$ s, and $1 \, M_\odot = 4.926 \times 10^{-6}$ s.

\section{General formalism for estimating statistical and systematic errors}
\label{sec:parameterestimation}

\subsection{A review of signal analysis}

This section briefly reviews the basic formulas of signal
analysis, partly to fix notation.  For a more complete
discussion, see Refs.\ \cite{1994PhRvD..49.2658C} or \cite{Wainstein:1962}. 

The output of $D$ detectors can be represented by the
vector $s^\alpha(t)$, $\alpha = 1,2,\ldots,D$. It is often convenient
to work with the Fourier transform of the signal; our convention is
\begin{equation}
\label{fourierT}
{\tilde s^\alpha}(f) \equiv \int_{-\infty}^{\infty}\, e^{2\pi i f t}s^\alpha(t)\, dt.
\end{equation}
The detector output $s^\alpha(t)$ is the sum of GWs
$h^\alpha(t)$ plus instrument noise $n^\alpha(t)$. 
We assume that the noise is stationary and Gaussian;
here ``stationarity'' means that the different Fourier components 
$\tilde n^\alpha(f)$ of the noise are uncorrelated; thus we have 
\begin{equation}
\label{pr}
\langle {\tilde n^\alpha}(f) \, {\tilde n^\beta}(f^\prime)^* \rangle = \frac{1}{2}
\delta(f - f^\prime) S_{h}^{\alpha \beta}(f),
\end{equation}
where ``$\langle \ \rangle$'' denotes averaging over realizations of the noise, and
$S_h^{\alpha \beta}(f)$ is the (one-sided) spectral density of the noise. 
By contrast, ``Gaussianity'' implies that each Fourier component has 
Gaussian probability distribution.  
Under these assumptions, we obtain a 
natural inner product on the vector space of signals:  
given two signals $g^{\alpha}(t)$
and $k^\alpha(t)$, we define $\left( {\bf g} \, | \, {\bf k} \right)$ by
\begin{equation}
\label{inner}
\left( {\bf g} \,|\, {\bf k} \right) = 4 \, \mathrm{Re} \int_0^{\infty} [S_h(f)^{-1}]_{\alpha \beta}\,
\tilde g^\alpha(f)^* \tilde k^\beta(f) \, df,
\end{equation}
where we sum over repeated indices.
Here and below we reduce index clutter by using boldface to represent multiple-detector \emph{abstract} signal vectors (e.g., $\bf k$ will denote both $k^{\alpha}(t)$ and $\tilde k^{\alpha}(f)$, which are the same vector represented in different bases).

In terms of the inner product \eqref{inner}, the probability for the 
noise to have some realization
${\bf n}_0$ is just
\begin{equation}
\label{pn0}
p({\bf n} = {\bf n}_0) \, \propto \, e^{- \left( {\bf n}_0\, |\, {\bf n}_0 \right) /2}.
\end{equation}
Thus, if the actual incident waveform is ${\bf h}$, the probability of measuring a signal ${\bf s}$ in the
detector output is proportional to 
$\exp\{-( \mathbf{s}-\mathbf{h} \, | \, \mathbf{s}-\mathbf{h})/2\}$.  Correspondingly, given a measured signal ${\bf s}$, the
gravitational waveform ${\bf h}$ that best fits the data (in the maximum-likelihood sense) is the one that
minimizes the quantity $( \mathbf{s}-\mathbf{h} \, | \, \mathbf{s}-\mathbf{h} )$.  This criterion has a simple geometric interpretation.
If $N$ is the number of source parameters,
the parametrized space of waveforms $\{\mathbf{h}(\theta^i)\}$ is an $N$-dimensional manifold embedded in the vector space
of all possible measured signals. Given the measured signal ${\bf s}$, the best-fit waveform $\mathbf{h}(\theta_\mathrm{bf}^i)$ is the point on the waveform manifold that lies closest to ${\bf s}$ according to the distance $( \mathbf{s}-\mathbf{h}(\theta_\mathrm{bf}^i) \, | \, \mathbf{s}-\mathbf{h}(\theta_\mathrm{bf}^i) )$.  The vector from ${\bf h}(\theta_\mathrm{bf}^i)$ to ${\bf s}$ is then normal to the
waveform manifold at $\theta_\mathrm{bf}^i$. It is also easily shown that the (matched-filtering) SNR of $\bf h$ is
just its norm $({\bf h}|{\bf h})^{1/2}$.
 
For a given incident GW, different realizations
of the noise will give rise to somewhat different best-fit
parameters.  For large SNR, however, the best-fit parameters will assume a
normal distribution centered on the correct values. 
Specifically, if $\theta_\mathrm{tr}^i$ are the true
physical parameters, and $\theta_\mathrm{bf}^i(\mathbf{n}) \equiv \theta_\mathrm{tr}^i + \Delta \theta^i(\mathbf{n})$ are the best-fit parameters, then for large SNR the parameter-estimation errors $\Delta \theta^i$ have
the normal distribution
\begin{equation}
\label{gauss}
p(\Delta \theta^i)=\,{\cal N} \, e^{-\frac{1}{2}\Gamma_{ij}\Delta \theta^i
\Delta \theta^j}.
\end{equation}
Here $\Gamma_{ij}$ is the \emph{Fisher information matrix} defined by
\begin{equation}
\label{sig}
\Gamma_{ij} \equiv \bigg( \frac{\partial {\bf h}}{\partial \theta^i}\, \bigg| \, 
\frac{\partial {\bf h}}{\partial \theta^j }\bigg),
\end{equation} 
and ${\cal N} = \sqrt{ {\rm det}(\Gamma / 2 \pi) }$ is the
appropriate normalization factor.  For large SNR, the 
variance--covariance matrix of the errors $\Delta \theta^i$ is given by
\begin{equation}
\label{bardx}
\left< {\Delta \theta^i} {\Delta \theta^j}
 \right>  = (\Gamma^{-1})^{ij} + O(\mathrm{SNR}^{-1})
\end{equation}
(see, e.g., Ref.\ \cite{Vallisneri:2007}).

%
%

\subsection{Theoretical parameter-estimation errors}
\label{sec:linearized}

We now give a simple geometric picture of theoretical errors due to inaccurate templates. Consider the following \emph{two} manifolds embedded in the vector space of possible data streams: a manifold of
the true GR waveforms $\{\mathbf{h}_\mathrm{GR}(\theta^i)\}$, and a
different manifold of approximate waveforms $\{\mathbf{h}_\mathrm{AP}(\theta^i)\}$ generated by
some approximation procedure.  Since the two manifolds can be
parametrized by the same physical coordinates (the masses of the
bodies, their spin vectors, etc.), there is a natural one-to-one
mapping between waveforms of the same coordinates on the two manifolds.
 
Now consider the measured detector data ${\bf s} = {\bf h}_\mathrm{GR}(\theta_\mathrm{tr}^i) + {\bf n}$,
where ${\bf n}$ is again the detector noise. Since in practice we have access only to the approximate waveforms,
the ${\bf h}_\mathrm{AP}(\theta_\mathrm{bf}^i)$ that lies closest to the data ${\bf s}$ 
yields our best guess for the binary's parameters: that is, we determine the best-fit $\theta_\mathrm{bf}^i$
by drawing the normal from ${\bf s}$ to the manifold of approximate
waveforms $\{\mathbf{h}_\mathrm{AP}(\theta^i)\}$.  The best-fit parameters $\theta_\mathrm{bf}^i$
satisfy
\begin{equation}
\label{toterr}
\Big( \partial_j \mathbf{h}_\mathrm{AP}(\theta_\mathrm{bf}) \Big|  
\mathbf{s} - \mathbf{h}_\mathrm{AP}(\theta_\mathrm{bf}) \Big) = 0.
\end{equation} 
Here and below we write $\theta$ for $\theta^i$ when the parameter index is not required for implicit summations or to identify a specific parameter. To first order in the error 
$\Delta\theta^i \equiv \theta_\mathrm{bf}^i - \theta_\mathrm{tr}^i$, 
we can write $\mathbf{h}_\mathrm{AP}(\theta_\mathrm{tr}) - \mathbf{h}_\mathrm{AP}(\theta_\mathrm{bf}) \approx - \Delta\theta^j
\partial_j \mathbf{h}_\mathrm{AP}(\theta_\mathrm{bf})$, so 
\newcommand{\hGR}{\mathbf{h}_\mathrm{GR}}
\newcommand{\hAP}{\mathbf{h}_\mathrm{AP}}
\newcommand{\tbf}{\theta_\mathrm{bf}}
\newcommand{\ttr}{\theta_\mathrm{tr}}
\begin{widetext}
\begin{equation}
\label{err2}
\begin{aligned}
\mathbf{s} - \hAP(\tbf) & = \mathbf{n} + \hGR(\ttr) - \hAP(\ttr) + \hAP(\ttr) - \hAP(\tbf) \\
& \approx \mathbf{n} + \hGR(\ttr) - \hAP(\ttr) - \Delta\theta^j \partial_j \hAP(\tbf).
\end{aligned}
\end{equation}
Defining $\Gamma_{ij}(\tbf) \equiv ( \partial_i \hAP(\tbf)|\partial_j \hAP(\tbf))$ and inserting Eq.\ \eqref{err2} into Eq.\ \eqref{toterr} we obtain
\begin{equation}
\Delta\theta^i 
 =  \Big(\Gamma^{-1}(\tbf)\Big)^{ij} \, \Big( \partial_j \hAP(\tbf) \Big| \, \mathbf{n} \Big)
 +  \Big(\Gamma^{-1}(\tbf)\Big)^{ij} \, \Big( \partial_j \hAP(\tbf) \Big| \, \hGR(\ttr) - \hAP(\ttr) \Big) \, .
\label{linerr1}
\end{equation}
\end{widetext}
Thus, to leading order,  $\Delta\theta^i$ is the sum of a \emph{statistical} contribution $\Delta_n \theta^i$ due
to noise and a \emph{theoretical} contribution $\Delta_\mathrm{th}\theta^i$ due to template inaccuracy, with 
\begin{align}
\Delta_n\theta^i  & =   \Big(\Gamma^{-1}(\tbf)\Big)^{ij} \Big( \partial_j \hAP(\tbf) \Big| \mathbf{n} \Big), \label{linerr2} \\
\Delta_\mathrm{th}\theta^i & =  \Big(\Gamma^{-1}(\tbf)\Big)^{ij} \Big( \partial_j \hAP(\tbf) \Big| \hGR(\ttr) - \hAP(\ttr)\Big). \label{linerr3}
\end{align}
If we happen to know both $\ttr$ and the noise realization $\mathbf{n}$, then these two equations can be used to determine $\tbf$; the experimental situation, however, is that we can determine that a certain $\hAP(\tbf)$ is the best-fit waveform for $\mathbf{s}$, but we are left to wonder about the error $\Delta \theta \equiv \tbf - \ttr$. From this viewpoint, Eq.\ \eqref{linerr2} leads to result \eqref{bardx} for the distribution of statistical error, which scales as $\mathrm{SNR}^{-1}$; by contrast, Equation \eqref{linerr3} cannot be used as it is written, since we do not know $\ttr$, which for a given $\tbf$ depends on the noise. To leading order we can however replace $\hGR(\ttr) - \hAP(\ttr)$ by $\hGR(\tbf) - \hAP(\tbf)$, yielding
\begin{equation}
\label{linerr3b}
\Delta_\mathrm{th}\theta^i =  \Big(\Gamma^{-1}(\tbf)\Big)^{ij} \Big( \partial_j \hAP(\tbf) \Big| \hGR(\tbf) - \hAP(\tbf)\Big),
\end{equation}
which is a noise-independent estimate of theoretical error \cite{cutler96unpublished,Flanagan:1998uq,Kocsis:2005fj}. Moreover, since the signal amplitude (and therefore the SNR) appears quadratically in both $\Gamma^{ij}$ and in the rightmost inner product of Eq.\ \eqref{linerr3b}, we see that $\Delta_\mathrm{th}\theta$ is independent of the SNR \footnote{This justifies the replacement made in obtaining Eq.\ \eqref{linerr3b}, which introduces a correction proportional to $O(\mathrm{SNR}^{-1})$.}, so it tends to be the limiting factor in extracting source-parameter information from the closest, strongest sources.
Because $\Delta_\mathrm{th}\theta$ does not decrease with
increasing SNR, nor would it ``average out'' if the
same event was measured by a large number of nearly identical detectors,
it can be considered a systematic error. This characterization, however, could be slightly misleading: 
while $\Delta_\mathrm{th} \theta$ is independent of noise (to lowest order), it 
\emph{does} depend on the true parameter values $\theta_\mathrm{tr}$ (e.g., the
size and direction errors in the sky location depend on the true location).

Going back to our geometric picture, we see that 
Eq.\ \eqref{linerr3b} identifies the signal $\hGR(\ttr)$ 
to which $\hAP(\tbf)$ is closest, as it would be given by solving Eq.\ \eqref{toterr} for $\ttr$ in the absence of noise, to leading order
in the error $\Delta_\mathrm{th} \theta$. In the next few sections, we
develop tools to estimate $\Delta_\mathrm{th} \theta$ that are more accurate than Eq.\ \eqref{linerr3b}. 

\subsection{A best-fit-by-ODE estimate of theoretical error}

\newcommand{\hla}{\mathbf{h}_\lambda}
\newcommand{\tla}{\theta(\lambda)}
\newcommand{\ttl}{\theta_\mathrm{tr}(\lambda)}
As discussed above, given the observed best-fit waveform $\hAP(\tbf)$, our goal is to find the  
true waveform $\hGR(\ttr)$ that is best fit by it, in the sense that
\begin{equation}
\Big( \hGR(\ttr) - \hAP(\tbf) \Big| \hGR(\ttr) - \hAP(\tbf) \Big)
\end{equation}
is minimized, and hence to find the associated parameter estimation errors $\Delta_\mathrm{th}\theta^i = 
\tbf^i - \ttr^i$.  This is a minimization problem, and it could be attacked in many ways (e.g,
by a straightforward grid search or by simulated annealing).  Standard methods, however, seem
likely to prove computationally unaffordable, especially because we wish to repeat the minimization for thousands of different $\tbf$
in order to explore the distribution of theoretical error over parameter space.
We have therefore developed an alternative minimization scheme that works rather well for our particular problem, and that
is computationally quite efficient.  The basic idea is to define a one-parameter family of
models $\hla(\theta)$ that interpolate smoothly between $\hAP(\theta) \equiv \mathbf{h}_{\lambda=0}(\theta)$ and $\hGR(\theta) \equiv \mathbf{h}_{\lambda=1}(\theta)$, and then to solve 
the minimization problem on the path from $\lambda = 0$, where we know trivially that $\ttr(0) = \tbf$, to $\lambda = 1$, where $\ttr(1)$ is the sought-after true-parameter vector for the observed $\tbf$, so that $\Delta_\mathrm{th} \theta = \tbf - \ttr(1)$. Namely, each $\ttr(\lambda)$ is the solution to
\begin{equation}
\label{ortho}
\Big(\hla(\ttl) - \hAP(\tbf) \, \Big| \,  \partial_j \hAP(\tbf)\Big) = 0,
\end{equation}
which, once again, implies that (at least locally) $\hAP(\tbf)$ provides the best fit to $\hla(\ttl)$.  Taking the derivative of Eq.~\eqref{ortho} with respect to $\lambda$ yields the ODE
\begin{multline}
\label{ode1}
\frac{d \theta^i}{d \lambda} \Big(\partial_i \hla(\ttl) \Big| \partial_j \hAP(\tbf) \Big)  =  \\
- \Big(\hla'(\ttl) \Big| \partial_j \hAP(\tbf) \Big),
\end{multline}
and hence
\begin{equation}
\label{ode2}
\frac{d \theta^i}{d \lambda} = -\big(\tilde{\Gamma}^{-1}(\ttl,\tbf) \big)^{ij} \Big(\hla'(\ttl) \Big| \partial_j \hAP(\tbf) \Big)
\end{equation}
where
\begin{equation}
\label{ode3}
\tilde{\Gamma}_{ij}(\ttl,\tbf) \equiv
\Big( \partial_i \hla(\ttl) \Big| \partial_j \hAP (\tbf) \Big).
\end{equation}
Here all partial derivatives are taken with respect to $\tbf$, while $\hla'(\theta) \equiv d \hla(\theta)/d \lambda$.
As anticipated above, the initial condition for Eq.\ \eqref{ode2} is just $\ttl(0) = \tbf$, and integrating to $\lambda = 1$ yields $\ttr$ and therefore $\Delta_\mathrm{th} \theta$ as a function of $\tbf$.

Of course, there is an infinite number of ways in which we could define the one-parameter family $\hla(\theta)$ that
interpolates between $\hAP(\theta)$ and $\hGR(\theta)$. Our choice is explained in detail in Sec.\ III, but a brief summary 
can be given here.  Ignoring modulations due to LISA's motion, the phase function $\Psi(f)$ has the PN expansion (up through 3.5 PN order)
\begin{equation}
\label{pn-psi}
\Psi(f) = A x + B + C x^{-5/3}\bigg[1 + \sum_{k =2}^7 \psi_k x^k \bigg] \, ,
\end{equation}
where $A, B, C$ and the $\psi_k$ ($k = 2, \ldots ,7$) are known constants, and $x \equiv (\pi M f)^{1/3}$.
The signal modulations imposed by LISA's motion
depend on the function $t(f)$, which represents the instant of time at which the GW frequency sweeps
through $f$, and which has the PN expansion
\begin{equation} 
t(f) = D + E x^{-8/3}\bigg[1 + \sum_{k =2}^7 \tau_k x^k \bigg] \, ,
\end{equation}
where $D, E$ and the $\tau_k$ are also known constants.
As discussed in the introduction, to construct $\hGR(\theta)$ we use these full expansions, while for $\hAP(\theta)$ we truncate them 
after $k=6$ (or equivalently, we re-set $\psi_7$ and $\tau_7$ to $0$). For the interpolating family $\hla(\theta)$, we 
make the simple replacement
\begin{equation}
\psi_7 \rightarrow \lambda \psi_7 \,, \quad \tau_7 \rightarrow \lambda \tau_7 \, .
\end{equation}
For this choice of $\hla(\theta)$, we find that our ODE [Eq.\ \eqref{ode2}] leads to a superb fits.
Defining the normalized inner product (sometimes called \emph{match}) between two waveforms as
\begin{equation}
\label{corr}
\mathrm{M}\big(\mathbf{g},\mathbf{k}\big) = \frac{\big(\mathbf{g}\big|\mathbf{k}\big)}{\big(\mathbf{g}\big|\mathbf{g}\big)^{1/2}\big(\mathbf{k}\big|\mathbf{k}\big)^{1/2}} \,,
\end{equation}
we generally find (see Sec.\ \ref{sec:results}) that the match between
the true and best-fit approximate waveforms is 
\begin{equation}
\label{overlap2}
\mathrm{M}\big(\hGR(\ttr),\hAP(\tbf)\big) > 0.9999.
\end{equation}
Incidentally, such high matches mean that it would be practically impossible to tell from the detector data alone that our theoretical waveforms were inaccurate, since the best approximate waveform would be a nearly perfect fit to the true waveform, and so would also fit the data equally well.

\subsection{An improved one-step estimate of theoretical error}

The ODE method described in the previous section suggests an improved version of the 
linearized estimate \eqref{linerr3b} derived in Sec.\ \ref{sec:linearized}.
We begin by writing each component $\tilde{h}_\lambda^\alpha(f)$ of the interpolating waveform 
as 
\begin{equation}
\label{ampphase}
\tilde h^{\alpha}_{\lambda}(f) =  A^{\alpha}_{\lambda}(f) e^{i \Psi^{\alpha}_{\lambda}(f)} \, ;
\end{equation}
Then 
\begin{equation}
\tilde{h}'^{\alpha}_{\lambda}(f) = \big[
A'^{\alpha}_{\lambda}(f) + i A^{\alpha}_{\lambda}
\Psi'^{\alpha}_{\lambda}(f)
\big]  e^{i \Psi^{\alpha}_{\lambda}(f)} \, ,
\end{equation}
so we can re-write Eq.~(\ref{ode2}) as
\begin{equation}
\label{ode2b}
\frac{d \theta^i}{d \lambda} =
-\underbrace{\big(\tilde{\Gamma}^{-1}\big)^{ij}}_{\text{Eq.\ \eqref{ode3}}}
\Big(
\underbrace{\big[\mathbf{A}'_{\lambda} + i \mathbf{A}_{\lambda} \bm{\Psi}'_{\lambda} \big] e^{i \Psi^{\alpha}_{\lambda}}}_{\text{at}\,\theta = \ttl}
\Big| 
\partial_j \hAP(\tbf)
\Big) \, .
\end{equation}
To estimate $\Delta_\mathrm{th} \theta^i$ in a single step (without solving the ODE), we approximate $\mathbf{A}'_{\lambda}(\ttl) $ and $\bm{\Psi}'_{\lambda}(\ttl)$, which are functions of the interpolating parameter $\lambda$, as constants:
\begin{equation}
\label{newrhs}
\begin{aligned}
\mathbf{A}'_{\lambda}(\ttl) &\approx \mathbf{A}_\mathrm{GR}(\tbf) - \mathbf{A}_\mathrm{AP}(\tbf) \equiv \Delta \mathbf{A} \, , \\
\bm{\Psi}'_{\lambda}(\ttl)  &\approx \bm{\Psi}_\mathrm{GR}(\tbf) - \bm{\Psi}_\mathrm{AP}(\tbf) \equiv \Delta \bm{\Psi} \, .
\end{aligned}
\end{equation}
Similarly, while $\tilde \Gamma_{ij} $ is also a function of $\lambda$, 
we find that in practice it is approximately constant; thus, to lowest order in a Taylor-series expansion about $\lambda = 0$, we replace it with
\be
\label{const-Gamma}
\tilde{\Gamma}_{ij}(\ttl,\tbf) \rightarrow \Gamma_{ij}(\tbf) \equiv \big( \partial_i \hAP(\tbf) \big| \partial_j \hAP(\tbf) \big) \, .
\ee
With these substitutions the right-hand side of Eq.~(\ref{ode2b}) becomes constant, and its integration trivial:
\be
\label{1-step}
\Delta_\mathrm{th}\theta^i \approx
\big(\Gamma^{-1}(\tbf) \big)^{ij}  \Big( \underbrace{\big[ \Delta {\bf A} + i {\bf A} \Delta \bm{\Psi} \big] e^{i \bm{\Psi}}}_{\text{at}\,\theta=\tbf} \Big| \,  \partial_j \hAP(\tbf) \Big).
\ee
Note this is quite similar to the linearized expression \eqref{linerr3b}, but with the replacement
\be
\label{one-step2}
\hGR - \hAP \equiv \Delta \mathbf{h} \rightarrow \big[
\Delta \mathbf{A} + i \mathbf{A} \Delta \bm{\Psi}
\big]  e^{i \bm{\Psi}}  \, .
\ee
Clearly, the two sides of Eq.\ \eqref{one-step2} agree up to terms of second order in $\Delta \mathbf{A}$ and $\Delta \bm{\Psi}$. However, our experience shows that Eq.\ \eqref{1-step} is more accurate  than Eq.\ \eqref{linerr3b} at determining theoretical errors. Why this is so is explained in the next subsection, where we present a second derivation of Eq.\ \eqref{1-step}.

\subsection{A second derivation of the improved one-step estimate}

As discussed in Sec.\ \ref{sec:results} below, we find in practice that while Eq.\ \eqref{linerr3b}
and our improved one-step estimate \eqref{1-step} agree for very small values 
of $\Delta_\mathrm{th}\theta$, Eq.\ \eqref{1-step} remains a good approximation for a much
larger range of $\Delta_\mathrm{th}\theta$. The basic reason is the following.  While in practice $\hAP(\tbf)$ 
turns out to be extremely close to $\hGR(\ttr)$, with matches exceeding $0.9999$, the difference
$\hAP(\tbf) - \hAP(\ttr)$ (which is the difference of two waveforms from the same family, but
evaluated at different points) is not well approximated by the first term in its Taylor expansion,
as assumed in deriving Eq.\ \eqref{err2} and therefore Eqs.\ \eqref{linerr3b}:
\be
\label{replace}
\hAP(\tbf)  - \hAP(\ttr) \not\approx \Delta \theta^j \partial_j \hAP(\tbf) \, . 
\ee 
Fortunately, the differences in both the amplitude and phase of the waveforms \emph{are}
individually well approximated by the linear terms in a Taylor series,
\begin{equation}
\label{Taylor}
\begin{aligned}
\mathbf{A}_\mathrm{AP}(\tbf)  - \mathbf{A}_\mathrm{AP}(\ttr) & \approx \Delta \theta^j \partial_j \mathbf{A}_\mathrm{AP}(\tbf) \, , \\
\bm{\Psi}_\mathrm{AP}(\tbf)  - \bm{\Psi}_\mathrm{AP}(\ttr)  & \approx \Delta \theta^j \partial_j \bm{\Psi}_\mathrm{AP}(\tbf) \, .
\end{aligned}
\end{equation}
For replacement \eqref{replace} to be reliable, the phase difference between the two waveforms should be much less than one
radian, $\Delta \theta^j \partial_j \bm{\Psi}_\mathrm{AP}(\tbf) \ll 1$; however, Eq.\ \eqref{Taylor} is reliable as long as $\Delta \theta^i \Delta \theta^j \partial_i \partial_j \bm{\Psi}_\mathrm{AP}(\tbf) \ll 1$, which is obviously a much less
restrictive condition \footnote{More generally, given a function $f(x) = e^{i g(x)}$, where $g(x)$ is
large and monotonically increasing, the linearized expansion $f(x) \approx f(x_0) + f'(x_0) (x-x_0)$
is obviously only reliable as long $f'(x_0) (x-x_0) \ll 1$.  It is better to approximate
$f(x)$ by $ e^{i [g(x_0) +  g'(x_0) (x-x_0)]} $, which is generally reliable as long as 
$g''(x_0) (x-x_0)^2 \ll 1$.}.

Given these considerations, we now provide an alternative derivation of our improved one-step formula \eqref{1-step}
along the same lines as our derivation of Eq.\ \eqref{linerr3b}. We begin again with Eq.\ \eqref{toterr}, but this time
for simplicity we neglect the noise term $\mathbf{n}$ that determines the statistical error. Defining
\begin{equation}
\label{defs1}
\begin{aligned}
\delta \mathbf{A} & \equiv \mathbf{A}_\mathrm{AP}(\tbf) - \mathbf{A}_\mathrm{GR}(\ttr) \, , \\
\delta \bm{\Psi} & \equiv \bm{\Psi}_\mathrm{AP}(\tbf) - \bm{\Psi}_\mathrm{GR}(\ttr) \, ,
\end{aligned}
\end{equation}
we can rewrite $\hGR(\ttr) - \hAP(\tbf)$ as
\newcommand{\AGR}{\mathbf{A}_\mathrm{GR}}
\newcommand{\AAP}{\mathbf{A}_\mathrm{AP}}
\newcommand{\PGR}{\bm{\Psi}_\mathrm{GR}}
\newcommand{\PAP}{\bm{\Psi}_\mathrm{AP}}
\begin{widetext}
\bea 
\AGR(\ttr) e^{i \PGR(\ttr)} - \AAP(\tbf) e^{i \PAP(\tbf)} &=&
\big(\AAP(\tbf) - \delta \mathbf{A} \big) e^{i ( \PAP(\tbf) - \delta \bm{\Psi} )}
- \AAP(\tbf) e^{i \PAP(\tbf)} \label{dAdpsi} \\
&\approx& - \big[\delta \mathbf{A} + i \delta \bm{\Psi} \AAP(\tbf) \big] e^{i \PAP(\tbf)} \, . \label{approx2}
\eea
In going from Eq.\ \eqref{dAdpsi} to Eq.\ \eqref{approx2}, we have neglected terms of order
$\delta \mathbf{A} \delta \bm{\Psi} $ and $\delta \bm{\Psi} \delta \bm{\Psi}$, which is justified because
$\delta \mathbf{A} / \AAP(\tbf) \ll 1$ and
$\delta \bm{\Psi} \ll 1$. Next, using Eqs.\ \eqref{Taylor} and Eqs.\ \eqref{newrhs} we can rewrite ${\bf \delta A}$ as
\newcommand{\dA}{\delta \mathbf{A}}
\newcommand{\dP}{\delta \bm{\Psi}}
$\dA \approx \Delta \theta^i \partial_i \AAP(\tbf) + \Delta \mathbf{A}$, and similarly
$\dP \approx \Delta \theta^i \partial_i \PAP(\tbf) + \Delta \bm{\Psi}$.
Using $\partial_i \mathbf{h} = [\partial_i \mathbf{A} + i \mathbf{A} \partial_i \bm{\Psi} ] e^{i \bm{\Psi}}$
along with Eq.\ \eqref{approx2}, we then have
\begin{equation}
\hGR(\ttr) - \hAP(\tbf) =  -\Delta \theta^i \partial_i \hAP(\tbf)
- \big[\Delta \mathbf{A} + i \mathbf{A} \Delta \bm{\Psi} \big] e^{i\bf{\Psi}_\mathrm{AP}} \, ,
\end{equation}
and plugging this expression into Eq.\ \eqref{toterr} while setting $\mathbf{n} = 0$, we again arrive at
\be\label{1-stepb}
\Delta_\mathrm{th}\theta^i \approx \big(\Gamma^{-1}(\tbf) \big)^{ij} \Big( \big[\Delta \mathbf{A} + i \mathbf{A}
\Delta \bm{\Psi} \big] e^{i \bm{\Psi}} \Big| \partial_j \hAP(\tbf) \Big) \, .
\ee
\end{widetext}

\subsection{A digression on parameter estimation in the face of theoretical uncertainties}

The method for inferring best-fit parameters represented by Eq.\ \eqref{toterr} -- simply use our best waveforms, as if they were exact, to infer the physical parameters of the source -- is of course a rather naive way of dealing with our theoretical limitations.
Since we know that our waveforms are not fully accurate, and especially since
we have some idea of where they fail most badly, we could certainly imagine
adopting other strategies. For instance, since we know that the PN approximation is less reliable for higher values of the PN expansion parameter $x \equiv (\pi M f)^{1/3}$, and 
therefore at higher frequencies, we could somehow modify the inner product \eqref{inner} to give less weight to the higher-frequency part of the waveforms. We take a first cut at this approach in Sec.\ \ref{sec:results}. 

An alternative, Bayesian approach to this problem would be to construct
a parametrized family of waveforms $\mathbf{h}(\theta,\alpha^K)$ that coincide up through all known PN orders, but differ at higher orders. Here the $\theta^i$ are again the source parameters, while the $\alpha^K$ parametrize our different models by fixing the unknown higher--PN-order terms (so in this context a ``model'' becomes just a rule to derive a waveform $\mathbf{h}$ for any given set of physical parameters $\theta^i$.) After assigning an \emph{a priori} probability $p(\alpha^K)$ to the different models (reflecting our belief about the degree to which the model matches GR), we could marginalize the Bayesian posterior probability for the $\theta^i$ over the models, as in 
\begin{equation}
p(\theta^i|\mathbf{s}) = \int p(\theta^i|\mathbf{s};\alpha^K) p(\alpha^K) d\alpha^K \, ,
\end{equation}
where $p(\theta^i|\mathbf{s}; \alpha^K)$ would be obtained by using $\mathbf{h}(\theta^i;\alpha^K)$ to build the likelihood.

Obviously the approach adopted in this paper is much less ambitious, but it seems a reasonable first pass at this problem; we expect that a more sophisticated approach would lead to smaller theoretical errors, so our results should provide a rough upper limit to the magnitude of $\Delta_\mathrm{th}\theta$.
If the naive error estimate is reassuringly small, there may be no need for a more sophisticated treatment.

\section{Model of the signal}
\label{sec:sigmodel}

To calculate $\Delta_\mathrm{th} \theta^i$ using the formalism developed in Sec.\ \ref{sec:parameterestimation}, we need to know the difference $\Delta \mathbf{h} \equiv \hGR(\tbf) - \hAP(\tbf)$ between the exact and approximate waveforms.
Since calculating $\hGR$ is currently beyond anyone's ability,
in practice we must content ourselves with estimating the rough magnitude of $\Delta_\mathrm{th} \theta$ from a reasonable estimate of $\Delta \mathbf{h}$.
Our basic strategy is to designate the 3.5PN restricted waveform as $\hGR$, and the corresponding 3PN restricted waveform for $\hAP$. We consider two variants of this:
in one version we write the phasing $\PAP$ through 3PN order, omitting the 3.5PN terms that go into $\PGR$; in the other, we augment $\PAP$ with the 3.5PN term of 
lowest order in the symmetric mass ratio $\eta$ (which, when small, approaches $M_2/M_1$).

This second version is motivated by the following considerations. We can clearly improve on the best PN waveforms by employing a hybrid approximation technique that expands the Einstein solution as a joint power series in both
$v/c$ and $\eta$.  The lowest-order terms in $\eta$ can be obtained to virtually arbitrary order in the PN expansion parameter, by means of a perturbative calculation for 
a test particle traveling on a geodesic around a BH.
This program was begun by Cutler and colleagues \cite{Cutler-Finn} and Poisson \cite{Poisson93}, for the case of quasi-circular orbits in Schwarzschild, and has since been extended to nearly circular, nearly equatorial orbits in Kerr spacetime~\cite{1995PThPh..93..307T,1996PhRvD..54.1439T,1995PhRvD..51.1646S} as well as to spinning particles in circular, equatorial Kerr orbits~\cite{1996PhRvD..54.3762T}; for a review, see Sasaki and Tagoshi~\cite{2003LRR.....6....6S}.
For the case of circular orbits around nonrotating BHs, this PN expansion is now known analytically
through 5.5PN order~\cite{Sasaki99};
that is, while binary-inspiral waveforms are fully known only through 3.5PN order,
the lowest-order pieces in $\eta$ are known analytically through 5.5PN order.
In constructing the best possible PN waveforms, it seems appropriate to include the extra information derived from BH perturbation theory. 
We shall refer to PN waveforms that have been supplemented in this way as ``hybrid'' waveforms. We do not know where this 
hybrid approach was first clearly spelled out, but it is implicit in Kidder, Will, and Wiseman \cite{1993PhRvD..47.3281K}, is discussed in Damour, Iyer, and Sathyaprakash \cite{Damour:1998lr}, and is used by Canitrot~\cite{Canitrot}.

As mentioned in Sec.\ \ref{sec:introduction}, because this paper represents a first pass at the problem of estimating theoretical errors for the case of MBHB inspirals, we make some simplifications: we work within the restricted PN approximation, and we neglect the spin-induced ``$\vec L \times \vec S$'' precession of the orbital plane, although we do include the lowest-order ``$\vec L \cdot \vec S$'' spin--orbit coupling term in the binary-inspiral phasing.
In addition, we model the LISA response to GWs using the
``low-frequency approximation'' of Cutler \cite{Cutler1998PhRvD..57.7089C}, and 
the LISA instrument noise using the analytic fit of Barack and Cutler \cite{2004PhRvD..70l2002B}.    
All these ingredients to our computation are briefly summarized in the following subsections.


\subsection{The LISA response to generic nonspinning-binary inspiral waveforms}
\label{sec:lisaresponse}

We describe the binary inspiral as an adiabatic sequence of quasicircular orbits in a fixed plane (since we are neglecting spin-induced precession) with normal $\hat{L}^a$.
Let $n^a$ be the unit vector that points from the detector to the source; in terms of the ecliptic colatitude and longitude angles $\bar{\theta}_S$ and $\bar{\phi}_S$, $n^a \equiv \{ \sin \bar{\theta}_S \cos \bar{\phi}_S, \sin \bar{\theta}_S \sin \bar{\phi}_S, \cos \bar{\theta}_S \}$ and likewise
$\hat{L}^a \equiv \{ \sin \bar{\theta}_L \cos \bar{\phi}_L, \sin \bar{\theta}_L \sin \bar{\phi}_L, \cos \bar{\theta}_L \}$.
Now let $p^a$ and $q^a$ be axes 
orthogonal to $n^a$, defined by
\bea
p^a &=& \epsilon^{abc} n_b \hat L_c \, , \\
q^a &\equiv&  -\epsilon^{abc} n_b p_c \, ; \label{pq}
\eea
we use these $p^a$ and $q^a$ to define the basis tensors for the GW polarization, 
\begin{equation}
H^+_{ab} = p_a p_b - q_a q_b  \,, \quad \ H^\times_{ab} = p_a q_b + q_a p_b \, .
\label{poldef}
\end{equation}  
Within the restricted PN approximation, the gravitational waveform
$h_{ab}(t)$ at the Solar System Barycenter can be written as
\begin{equation}\label{generalwave}
h_{ab}(t) = A_+(t) H^+_{ab} \cos \Phi(t) +  A_\times(t) H^\times_{ab} \sin\Phi(t)\,
\end{equation}
where $\Phi(t)$ is the time-domain PN phasing, and the amplitudes $A_+$ and $A_\times$ are given by
\begin{equation}
\big\{ A_+(t) , A_\times(t) \big\} = \frac{2 M_1 M_2}{r(t) D} \big\{ \alpha_+ , \alpha_{\times} \big\} 
\end{equation}
with $M_1$, $M_2$ the two masses, $r(t)$ the instantaneous orbital separation, $D$ the distance to the binary, and
\begin{equation}
\label{amps2}
\big\{\alpha_+,\alpha_{\times}\big\} = \big\{ \big[1 + (\hat L^a n_a)^2 \big], - 2\hat L^a n_a \big\} \, .
\end{equation}

For the MBHB inspirals considered here, most of the LISA SNR accumulates at frequencies
$f < 10\,$mHz, so it is adequate to use the low-frequency approximation to the
LISA response functions derived by Cutler \cite{Cutler1998PhRvD..57.7089C}.
In this approximation, the LISA science data consist of two independent (i.e., uncorrelated-noise) channels I and II \footnote{More generally, 
the LISA science data can be taken to consist of three 
Time-Delay Interferometry (TDI) channels with uncorrelated instrument noises, known as $A$, $E$, and $T$ \cite{2002PhRvD..66l2002P}.
However, only two of these ($A$ and $E$) have good sensitivity at frequencies below
$\sim 20\,$mHz, where most of the LISA SNR is accumulated for MBHB inspirals. 
At these lower frequencies, the $A$ and $E$ channels are basically equivalent to 
the channels I and II of Ref.\ \cite{Cutler1998PhRvD..57.7089C}, which are also used in this paper.}. The LISA channel-I response $h_\mathrm{I}(t)$ is
\begin{eqnarray}
h_\mathrm{I}(t) &=& \frac{\sqrt{3}}{2} A_+ F_\mathrm{I}^+(\theta_S,\phi_S,\psi_S) \cos \big[\Phi(t) + \varphi_D(t) \big]
\nonumber \\
\mbox{} &+& \frac{\sqrt{3}}{2} A_{\times} F_\mathrm{I}^{\times}(\theta_S,\phi_S,\psi_S) \sin \big[\Phi(t) + \varphi_D(t) \big] \,, \label{ht}
\end{eqnarray}
where the \emph{Doppler phase}
\be\label{Dphase}
\varphi_D(t) = \frac{d\Phi(t)}{dt} \,  \vec R(t) \cdot {\hat n}
\ee
is the difference between the phase of the wavefront at the LISA detector [located at $\vec R(t)$]
and at the Solar System Barycenter, and where $F_\mathrm{I}^+$ and $F_\mathrm{I}^\times$ are the detector beam-pattern coefficients \cite{Thorne:1987lr}
\begin{multline}
\label{Fpc}
F_\mathrm{I}^{+,\times}(\theta_S,\phi_S,\psi_S) = {\textstyle \frac{1}{2}}(1 + \cos^2 \theta_S) \cos 2\phi_S \cos
2\psi_S \\ \mp \cos \theta_S \sin 2\phi_S \sin 2\psi_S\, .
\end{multline}
Here $\theta_S(t)$, $\phi_S(t)$ and $\psi_S(t)$ (without overbars) are the source sky-location and polarization angles
with respect to a frame that rotates with the LISA detector, so they vary on a 1-year timescale.  Of course, these time-varying angles
depend on the fixed angles $\bar\theta_S$, $\bar\phi_S$, $\bar\theta_L$, and $\bar\phi_L$, although the relations between them
are slightly complicated. We refer the reader to section III B of Cutler \cite{Cutler1998PhRvD..57.7089C} for explicit expressions for $\theta_S(t)$, $\phi_S(t)$ and $\psi_S(t)$.
 
It is convenient to rewrite the signal (\ref{ht}) in the conventional amplitude--phase form:
\begin{equation}
h_\mathrm{I}(t)= \frac{\sqrt{3} M_1 M_2}{r(t) D} \, {\cal A}_\mathrm{I}(t) \cos [\Phi(t) + \varphi_{p,\mathrm{I}}(t) 
+ \varphi_D(t)]\,,
\label{h-amp-phase}
\end{equation}
where 
${\cal A}_\mathrm{I}(t)$ and $\varphi_{p,\mathrm{I}}(t)$ are given by
\begin{gather}
{\cal A}_\mathrm{I}(t) =
\Big(\alpha_+^2 {F_\mathrm{I}^+}^2(t)  + \alpha_\times^2 {\mathrm{F}_I^\times}^2(t) \Big)^{1/2} \,, 
\label{amp} \\ 
\varphi_{p,\mathrm{I}}(t) = \tan^{-1}\left(\frac{-\alpha_\times F_\mathrm{I}^\times(t)}{\alpha_+ F_\mathrm{I}^+(t)}\right)\,.
\label{pphase}
\end{gather}

Equations \eqref{ht}--\eqref{pphase} determine $h_\mathrm{I}(t)$ completely, while the prescription for $h_\mathrm{II}(t)$ is the same, 
except that the antenna patterns $F_\mathrm{I}^{+,\times}$ are replaced by
$F_\mathrm{II}^{+,\times}(\theta_S,\phi_S,\psi_S) = F_\mathrm{I}^{+,\times}(\theta_S,\phi_S - \pi/4,\psi_S)$.

\subsection{Post-Newtonian templates in the stationary-phase approximation}
\label{sec:pnspa}

In this paper we work with the Fourier transform $\tilde h_{\alpha}(f)$ of the LISA response (where $\alpha = \mathrm{I}, \mathrm{II}$), which
we approximate using the stationary phase approximation (SPA) \cite{Cutler1998PhRvD..57.7089C}: for $f>0$,
%
\begin{multline}
\tilde h_\alpha(f) = {\cal A}_\alpha\big(t(f)\big)  \sqrt{\frac{5}{128}} M_c^{5/6} f^{-7/6} \\ \times e^{i\bigl(2\pi (f - f_0) t_0 + \Lambda(f)  -  \Lambda(f_0) + \Lambda_0 \bigr)}
\label{stat}
\end{multline}
with
\be
\Lambda(f)  =  \Psi(f)  - \varphi_{p,\alpha}(t(f)) -  \varphi_D(t(f)) \, .
\label{stat2}
\ee
Here $M_c$ is the binary's chirp mass, which is related to the total mass $M \equiv M_1 + M_2$ and the symmetric mass ratio $\eta \equiv M_1 M_2/M^2$ by $M_c \equiv M \eta^{3/5}$; also, $\Lambda_0$ and $t_0$ are the phase and time at the fiducial frequency $f_0$, and $t = t(f)$ is the instant at which the GW frequency sweeps through the value $f$, as given below in Eqs.\ \eqref{pnt0} and \eqref{pnt}. Since $h_\alpha(t)$ is real, $\tilde{h}_\alpha(-|f|) = \tilde{h}^*_\alpha(|f|)$.

The main missing ingredient to Eqs.\ \eqref{stat} and \eqref{stat2} is the SPA PN phasing, given up through 3.5PN order in Refs.\ \cite{2001PhRvD..63d4023D,2002PhRvD..66b7502D}:
\begin{equation}
\label{pnPsi}
\Psi(f) =
\frac{3}{4}\bigl(8 \pi M_c f
\bigr)^{-5/3} \Big[1 + \sum_{k=2}^7 \psi_k x^k + \sum_{k=5}^6 \bar\psi_k x^k \log(x) \Big],\end{equation}
with the PN expansion parameter $x(f) \equiv (\pi M f)^{1/3}$, and
\begin{eqnarray}
\psi_2&=&\frac{20}{9}\left( \frac{743}{336} + \frac{11}{4}\eta
\right),\label{eq:beta2}\\
\psi_{3}&=& -16\pi + 4\beta\,, \label{eq:beta3}\\
\psi_4&=&10\left( \frac{3058673}{1016064} + \frac{5429\,
}{1008}\,\eta + \frac{617}{144}\,\eta^2 \right),\\
\bar\psi_5&=&\pi\left( \frac{38645 }{252}\,- \frac{65}{3}\eta\right),
\end{eqnarray}
\begin{eqnarray}
\psi_{6}&=&\left(\frac{11583231236531}{4694215680} - \frac{640\,{\pi
}^2}{3} - 
\frac{6848\,\gamma }{21}\right) \nonumber \\
&+& \!\!\eta \left( - \frac{15335597827}{3048192} + \frac{2255\,{\pi }^2}{12} 
- \frac{1760\,\theta }{3} +\frac{12320\,\delta }{9} \right)\nonumber\\
&+& \!\!\frac{76055}{1728}\eta^2-\frac{127825}{1296}\eta^3-\frac{6848}{21}
\log(4)\,,\\
\bar\psi_6 &=& -\frac{6848}{21}\,, \\
\psi_7 &=&\pi\left(\frac{77096675 }{254016} + \frac{378515
}{1512}\,\eta - \frac{74045}{756}\,\eta^2\right). \label{eq:betas}
\end{eqnarray}
In these expressions, $\gamma \equiv 0.57721\cdots$ is the Euler--Mascheroni constant, 
$\delta \equiv -1987/3080$, and $\theta = -11831/9240$.
Note that $\psi_5$ is not listed because it can be simply reabsorbed into a re-definition of $\Lambda_0$. In Eq.\ \eqref{eq:beta3} for $\psi_3$, $\beta$ is the 1.5PN spin--orbit phasing term defined in Ref.\ \cite{1994PhRvD..49.2658C} 
($\beta$ is conserved only approximately by the 1.5PN equations of motion, but in our model we treat it as a constant). Because we are not modeling spin--orbit \emph{precession} effects, we have chosen to include only the lowest-order spin effects in $\Psi(f)$; in
particular, we have not included any spin--spin terms, or any spin--orbit terms of order
higher than 1.5PN.

To 3.5PN order, $t(f)$ is given by
\begin{equation}
t(f) = t_0 + \tau(f) - \tau(f_0) \, ,
\label{pnt0}
\end{equation}
\begin{multline}
\tau(f) = - 5(8\pi f)^{-8/3} {M_c}^{-5/3} \\
\times \bigg[1 + \sum_{k=2}^7 \tau_k x^k + \sum_{k=5}^6 \bar\tau_k x^k \log(x) \bigg] \, ,
\label{pnt}
\end{multline}
%
where $\tau_k = \frac{5-k}{5}\psi_k$ for $k < 5$,
$\tau_5 =  -\frac{1}{5}\bar\psi_5$, 
$\tau_6 = -\frac{1}{5}(\psi_6 + \bar\psi_6)$, 
$\bar\tau_6 = -\frac{1}{5}\bar\psi_6$, and 
$\tau_7 =  -\frac{2}{5}\psi_7$.

We truncate the waveform of Eq.\ \eqref{stat} above a frequency corresponding to either $r = 6M$ or (to explore the theoretical error due to the last phase of inspiral) $r = 9M$, and below an initial frequency chosen by solving Eq.\ \eqref{pnt} to enforce a signal of a certain fiducial length (typically one year). 
Thus, our signal model includes just the inspiral waveform, and not
the final merger and ringdown. This is in part because of our current ignorance about the final merger: 
NR computations are now producing robust merger waveforms for nonrotating BHs \cite{Baker:2007fk}, and the case of generic spins is beginning to be explored 
\cite{Campanelli:2006lr,Campanelli:2007fk,Herrmann:2007qy,Brugmann:2007uq,Pollney:2007fj},
but the relevant parameter space is very large, and much work is still needed.
In addition, in the particular problem considered in this paper, most of the interesting information about binaries (and especially about their sky locations) is presumably encoded in their phase evolution during the slow inspiral, 
so one may hope that neglecting the information in the merger and ringdown
does not strongly impact parameter estimation, even in cases where the latter phases dominate
the total SNR. Establishing whether this is true certainly deserves careful investigation in the future.

Overall, $\tilde h_\alpha(f)$ depends on ten physical parameters:
the mass parameters $M_c$ and $\eta$, the spin--orbit parameter $\beta$,
the phase $\Lambda_0$ and time $t_0$ at the fiducial frequency $f_0$, 
the ecliptic sky-position angles $\bar{\theta}_S$ and $\bar{\phi}_S$ and orbital--angular-momentum angles $\bar{\theta}_L$ and $\bar{\phi}_L$ (which together determine the LISA-frame angles $\theta_S$, $\phi_S$, and $\psi_S$), and the distance $D$.
Note that for simplicity we treat the background spacetime as
flat, rather than Friedman--Lemaitre--Robertson--Walker.  Accounting for cosmological effects simply requires the translation $M_i \rightarrow M_i(1+z)$ and $D \rightarrow D_L$, where $D_L$ is the luminosity distance~\cite{1993PhRvD..48.4738M}.

\subsection{The fiducial, approximated, and interpolating waveform families}
\label{sec:families}

Our fiducial ($\hGR$), approximated ($\hAP$), and interpolating ($\hla$) template families are based on the LISA response model and SPA PN waveforms outlined in Secs.\ \ref{sec:lisaresponse} and \ref{sec:pnspa}.
\begin{itemize}
\item To build $\hGR$, we evaluate $\Psi(f)$ and $t(f)$ from Eqs.\ \eqref{pnPsi}--\eqref{pnt}, including all terms given there (through 3.5PN).
\item We build two distinct versions of $\hAP$: a \emph{straight} version that is the same as $\hGR$, except that we set $\psi_7$ and $\tau_7$ to zero by hand; and a \emph{hybrid} version where we set $\psi_7$ and $\tau_7$ to their $O(\eta^0)$ terms alone,
\begin{equation}
\psi_7 \rightarrow \pi \left(\frac{77096675}{254016}\right),
\end{equation}
and $\tau_7 \rightarrow -(2/5) \times (\text{the new $\psi_7$})$.

(Recall the motivation for the hybrid version: we assume that terms of $\psi_j$ and $\tau_j$
of lowest order in $\eta$ are known from BH perturbation theory, so $\hAP$ must converge to $\hGR$ as $\eta \rightarrow 0$.)
\item We also use two versions of the interpolating family $\hla$. The straight version 
is the same as $\hGR$, except that we modulate the strength of the 3.5PN terms with the replacement $\psi_7 \rightarrow \lambda \psi_7$ and $\tau_7 \rightarrow \lambda \tau_7$; 
clearly this family coincides with the straight $\hAP$ when $\lambda = 0$, and
with $\hGR$ when $\lambda = 1$.

The hybrid version is obtained with the replacement
\begin{equation}
\psi_7 \rightarrow \pi\left(\frac{77096675}{254016}\right) + \lambda \bigg[\pi \left( \frac{378515}{1512}\,\eta - \frac{74045}{756}\,\eta^2\right)\bigg],
\end{equation}
and $\tau_7 \rightarrow -(2/5) \times (\text{the new $\psi_7$})$, so that again $\hla \equiv \hAP$ for $\lambda = 0$ and $\hla \equiv \hGR$ for $\lambda = 1$.
\end{itemize}

\subsection{LISA noise model}
\label{NoiseModel}

We assume that the LISA noises in channels I and II are uncorrelated,
so the inner product between the signals ${\bf g} \equiv \big(g_\mathrm{I}(t), g_\mathrm{II}(t)\big)$ and ${\bf k} \equiv \big({k}_\mathrm{I}(t), k_\mathrm{II}(t)\big)$ is just  
\begin{equation}
\big( {\bf g} \,\big|\, {\bf k} \big)
\equiv 4 \, \mathrm{Re} \sum_{\alpha=\mathrm{I},\mathrm{II}} \int_0^{\infty}\frac{\tilde g_\alpha^*(f)
\tilde k_\alpha(f)}{S_h^{\alpha}(f)}\,df \, .
\end{equation}
We also also assume that $S^\mathrm{I}_h(f)$ and $S^\mathrm{II}_h(f)$ are exactly equal, and given by the fitting function $S_h(f)$ of Barack and Cutler \cite{2004PhRvD..70l2002B}.
(Note however that the $S_h(f)$ used here is actually $3/20$ times the expression quoted in Ref.\ \cite{2004PhRvD..70l2002B}.  This is because Ref.\ \cite{2004PhRvD..70l2002B} uses the sky-averaging convention, but this paper does not.)
The LISA noise has three main components:
instrument noise, confusion noise from short-period galactic white-dwarf (WD) binaries, and 
confusion noise from extragalactic WD binaries.  (We neglect confusion noise from unresolved
extreme-mass-ratio inspirals, since its magnitude is quite uncertain, and
since it is likely to be dominated by either instrument noise or WD confusion noise at all frequencies \cite{2004PhRvD..70l2002B}.)

For the LISA instrument noise we adopt the fitting function of 
Finn and Thorne \cite{2000PhRvD..62l4021F}, which is based on the noise budgets specified in the LISA Pre-Phase A Report \cite{Team:1998lr}:
\begin{multline}
\label{noise-Sum}
S^{\rm inst}_h(f) = \Big( 9.18 \times 10^{-52} (f/\mathrm{Hz})^{-4} + 1.59\times 10^{-41}
\\ + 9.18 \times 10^{-38} (f/\mathrm{Hz})^{2} \Big) {\rm Hz}^{-1}.
\end{multline}
Next we turn to confusion noise from galactic and extragalactic WD binaries
(GWDs and EWDs, respectively).
Any isotropic background of individually unresolvable GW sources represents
(for the purpose of analyzing {\it other} sources) a noise source with
spectral density \cite{2004PhRvD..69l2004A}
\be
\label{ShOm}
S^{\rm bgd}_h(f) = \, \frac{3}{5 \pi} f^{-3}  \frac{d\rho_{\rm GW}}{d(\log f)} \, ;
\ee
estimates of $d\rho_{\rm GW}/d\log f$ from the GWD and EWD backgrounds \cite{2003MNRAS.346.1197F,2001A&A...375..890N} yield the spectral densities
\begin{eqnarray}
 S^{\rm GWD}_h(f) &=& 2.1\times10^{-45}\,\left(\frac{f}{{\rm Hz}}\right)^{-7/3}
{\rm Hz}^{-1}, \\
S_h^{\rm EWD}(f) &=&
4.2 \times 10^{-47} \left(\frac{f}{{\rm Hz}}\right)^{-7/3}
{\rm Hz}^{-1}.
\end{eqnarray}
\noindent
Essentially none of the EWDs can be individually resolved and fit out, so to a first approximation 
one can treat $S_h^{\rm EWD}(f)$ as if it were just another source of instrument noise.
The situation is different for the GWDs.
While this background is actually larger than the LISA instrument noise between $\sim 10^{-4}$ and $10^{-2}$ Hz, at frequencies
$f \agt 2 \times 10^{-3}\,$Hz  the galactic sources
are sufficiently sparse, in frequency space, that one expects to be able
to fit them out of the data.  This process effectively reduces the
amount of information about any \emph{other} signal ${\bf h}$ in the data---this is because 
the piece of $\mathbf{h}$ that lies in the tangent space to the signal manifold that describes the resolvable WD binaries gets fitted out as part of the WD background \cite{2006PhRvD..73d2001C}.

Consider a frequency band of width $\Delta f$, much smaller than the 
LISA bandwidth, but much larger than $1/T_\mathrm{tot}$, where $T_\mathrm{tot}$ is LISA's data-collecting lifetime.
The number of independent data points collected in that band is $4 T_\mathrm{tot} \Delta f$, where the factor $4$ is the product of 2 data channels (I and II) times 2 real numbers (or one complex number) per discrete 
frequency bin.
The dimensionality of the subspace that is given up to the WDB background is  $7 (dN/df) \Delta f$, where
$dN/df$ is the density of GWDs in frequency space and $7$ is the number of parameters
required to specify each  GWD.  Thus the fraction of the information in ${\bf h}$ that is effectively lost 
to the background is $(7/4) \, dN/df \, T^{-1}_\mathrm{tot}$, and $S_h(f)$ is effectively multiplied
by $[1 - (7/4) \, dN/df \, T^{-1}_\mathrm{tot}]^{-1}$ \cite{2006PhRvD..73d2001C}. This simplistic counting clearly breaks down as
$(7/4) \, dN/df \, T^{-1}_\mathrm{tot} \rightarrow 1$, since the total effective noise should not exceed
the sum of all the individual noises. Lacking a more sophisticated treatment of this transition, 
we adopt the following fitting function as 
our (admittedly crude) estimate of LISA's total effective noise density $S^{\rm eff}_h(f)$:
\begin{equation}
\label{Stot}
\begin{aligned}
S^{\rm eff}_h(f) = {\rm min}
\Big\{ & \left[S_h^{\rm inst}(f) + S_h^{\rm EWD}(f) \right]
\exp(\kappa T^{-1} dN/df), \\
& S_h^{\rm inst}(f) + {\cal S}_h^{\rm GWD}(f) + S_h^{\rm EWD}(f) \Big\}.
\end{aligned}
\end{equation}
For $dN/df$ we adopt the estimate \cite{2002MNRAS.331..805H}
\begin{equation}\label{eq:dNdf}
\frac{dN}{df} = 2\times10^{-3}\,{\rm Hz}^{-1}\left(\frac{1\,{\rm Hz}}{f}\right)^{11/3}.
\end{equation}
In the paragraph above Eq.\ \eqref{Stot} we have given an argument suggesting that $\kappa = 7/4$ (assuming that both channels I and II are operational),
but this really represents the best one could do. Partly to be consistent with
earlier papers, we account for suboptimal WD subtraction
by adopting a more conservative estimate for $\kappa$
given by $\kappa T_\mathrm{tot}^{-1} = 1.5/{\rm yr}$.  Because $dN/df$ falls so steeply with
increasing $f$, had we (very optimistically) taken $\kappa T_{tot}^{-1} = 7/(20{\rm yr})$,
the steep drop off in $S^{\rm eff}_h(f)$ at $f \sim 2$ mHz would merely have been
shifted a factor $\sim 1.5$ towards lower $f$.

\section{Results}
\label{sec:results}

To test our formalism numerically, we have developed MATLAB code to do the following: a) generate the gravitational waveforms, LISA response, and LISA noise spectrum described in Sec.\ \ref{sec:sigmodel}; b) compute noise inner products according to Eq.\ \eqref{inner}; and c) implement the ODE [Eqs.\ \eqref{ode2}--\eqref{ode3}] and improved one-step [Eq.\ \eqref{1-step}] formulas for the theoretical error. We use analytical derivatives of the signals with respect to all parameters except the PN-interpolating parameter $\lambda$ and the sky-position and $\vec{L}$ angles, for which we use central finite differences with displacements of $10^{-7}$ and $2 \pi \times 10^{-5}$, respectively (the waveforms and LISA response change very smoothly with these parameters, so this approximation is adequate). In addition, the following details about our implementation are worth pointing out:
\begin{itemize}
\item The PN-interpolation parameter $\lambda$ enters the LISA-response expressions of Sec.\ \ref{sec:lisaresponse} through $t(f)$ [Eqs.\ \eqref{pnt0} and \eqref{pnt}]; we include this dependence when computing numerical derivatives.
\item In principle, the derivative $\partial \mathbf{h} / \partial \log M_c$ includes terms of the form $(\partial \mathbf{h}/\partial F_\mathrm{I, II}) \times (\partial F_\mathrm{I, II}/\partial t) \times (\partial t/\partial \log M_c)$. However, these contributions are quite time-consuming to compute, and they are also $\sim 10^5$ times smaller than the dominant terms of the form $(\partial \mathbf{h}/\partial \Psi) \times ( \partial \psi / \partial \log M_c)$. Therefore we neglect the former, as well as similar terms in the derivatives $\partial \mathbf{h} / \partial \eta$ and $\partial \mathbf{h} /\partial \beta$.
\item We choose the Nyquist frequency and frequency spacing used to represent waveforms in the frequency domain in such a way to avoid both frequency aliasing and time wrapping.
\item The artificial truncation of waveforms at GW frequencies corresponding to $r = 6M$ (or $9M$) induces a mildly pathological behavior in the inner products of signals from systems with different total masses; we deal with this problem by truncating all waveforms at the $r = 6M$ (or $9M$) GW frequency of the $\hAP(\tbf)$ for which we are estimating theoretical error. This approximation is justified because the frequency-domain truncation does not represent any physical feature of the true waveforms, but stems instead from our ignorance about the late stages of their inspiral. Furthermore, the truncation does not affect the information carried by the signals at lower frequencies, and while it \emph{does} exclude some high-frequency information, it is information that we are ill-equipped to interpret anyway.
\item Last, we set the fiducial frequency $f_0$ to $90\%$ of the truncation frequency; since most of the total SNR for the systems studied in this paper comes from the last stages of inspiral, this has the effect of reducing the covariance in the joint estimation of $\Lambda_0$ and $t_0$ [see Eq.\ \eqref{stat}], and therefore of increasing the robustness of our framework (and especially of the one-step formula). 
\end{itemize}
In the remainder of this section, we test the accuracy of the improved one-step formula against the solution of the ODE, and we survey theoretical errors for MBHB systems with component masses between $10^4$ and $10^7 M_\odot$.

\subsection{Comparison of the ODE and improved one-step formulas}

To evaluate the accuracy of the improved one-step formula against the ODE formula, we select a representative system with $M_1 = 0.5 \times 10^6 M_\odot$, $M_2 = 10^6 M_\odot$, and $\beta = 0$, and we compute expected theoretical errors for 600 random choices of the sky-position and $\vec{L}$ angles, uniformly distributed on their respective spheres. (Recall also that the theoretical errors are independent of SNR, so there is no need to specify the latter.) All signals are created with a duration of one year (i.e., they begin a year before the system reaches the truncation frequency); the LISA orbits are chosen according the conventions of Ref.\ \cite{Cutler1998PhRvD..57.7089C}, and time-shifted so that the LISA position and orientation are the same for all binaries at the time when they achieve their respective fiducial frequency $f_0$.

\begin{turnpage}
\begin{figure}
\includegraphics{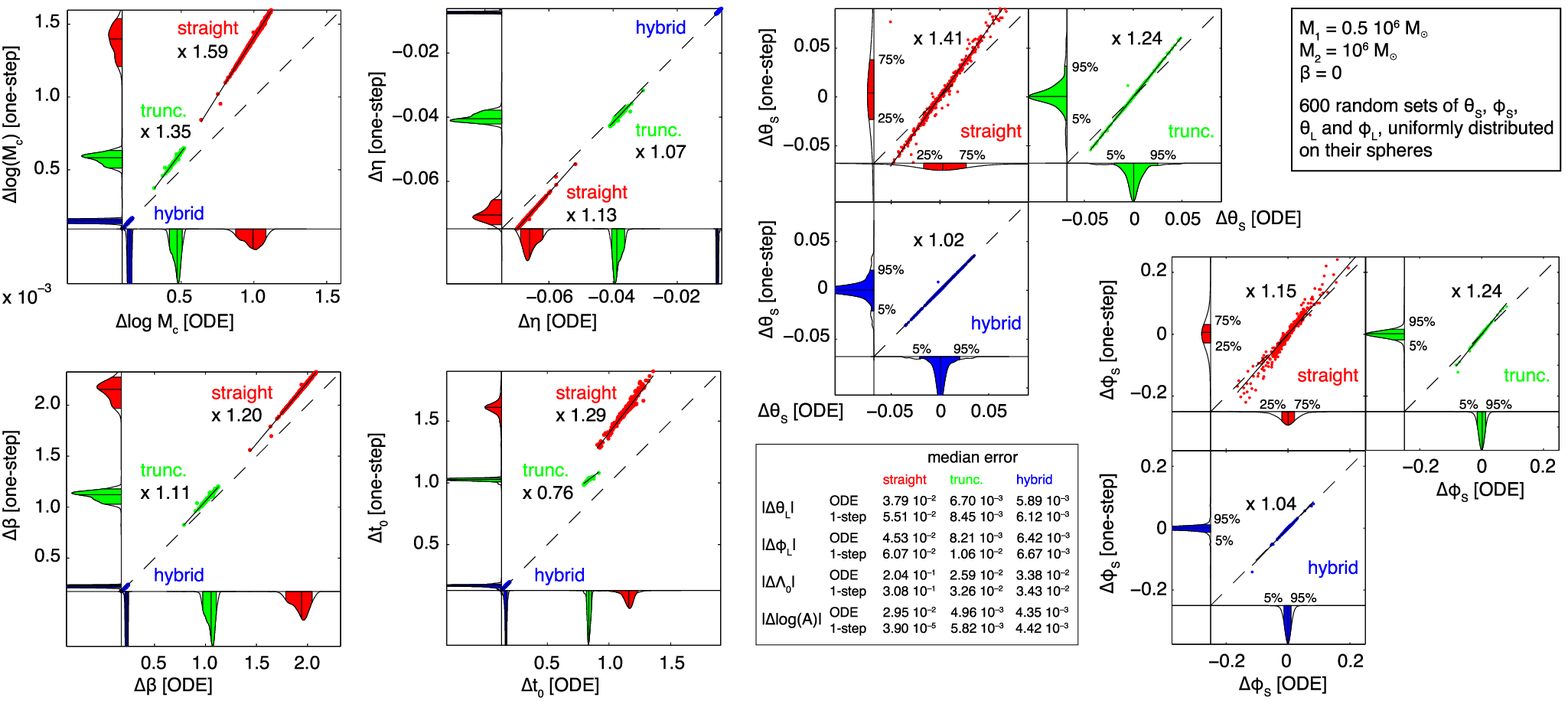}
\caption{Scatter plots of the ODE theoretical errors (horizontal axes) against the improved--one-step theoretical errors, for MBHB waveforms with $M_1 = 0.5 \times 10^6 M_\odot$, $M_2 = 10^6 M_\odot$, and $\beta = 0$, over populations of 600 sets of sky-position and $\vec{L}$ angles. The clusters labeled ``straight,'' ''trunc.,'' and ``hybrid'' refer to errors for the three waveform variants considered in this paper (truncated at $r = 6M$, at $r = 9M$, and with hybrid phasing). The margins of each plot show the one-variable distribution of the two types of error, with vertical bars marking the location of the median, and of the 5\% and 95\% quantiles (except where stated differently); these plots were normalized arbitrarily, but consistently for each parameter.
The lines drawn through the ``straight'' and ``trunc.'' clusters represent least-squares fits to $\Delta \theta_\mathrm{one\mbox{-}step}(\Delta \theta_\mathrm{ODE})$, with the correspondent linear coefficient indicated nearby (by comparison, the dashed lines are the locus of $\Delta \theta_\mathrm{one\mbox{-}step} = \Delta \theta_\mathrm{ODE}$).
Scatter plots are not shown for parameters $\bar\theta_L$, $\bar\phi_L$, $\Lambda_0$, and $A$ for which the one-step formula can be farther off the ODE prediction (although medians are still close, as shown in the in-figure table).
\label{fig:onestepvsode}}
\end{figure}
\end{turnpage}

The ODE and one-step errors are compared in the scatter plots of Fig.\ \ref{fig:onestepvsode}, where the clusters labeled ``straight,'' ``trunc.,'' and ``hybrid'' refer, respectively, to errors computed for our original waveforms truncated at $r = 6M$, for waveforms truncated at $r = 9M$, and for waveforms with the hybrid phasing described in Sec.\ \ref{sec:families} (and again truncated at $r = 6M$). The margins of each scatter plot display the single-variable distribution of the two errors over the populations of sky positions and $\vec{L}$ orientations. We see that for the ``intrinsic'' parameters $M_c$, $\eta$, $\beta$, for the sky-location angles, and for the time $t_0$, the one-step formula provides a reasonable approximation for the ODE result, which improves in going from the straight signal model to the truncated model, and (even more so) to the hybrid model.

In fact, the one-step errors appear to deviate from the ODE results by almost-constant multiplicative factors (approximated in the figure by least-squares--fitting the clusters to lines), which indicates that the error incurred in reducing the ODE to a single step is roughly independent of sky-position and $\vec{L}$ angles. Unfortunately, at present we do not have a way to predict the corrective factors from the parameters of the system. The behavior of the one-step formula is more erratic for the remaining parameters (the $\vec{L}$ angles $\bar\theta_L$ and $\bar\phi_L$, the phase $\Lambda_0$, and the amplitude $A$), with many outliers that are several times the maximum reported ODE error, which makes it hard to prepare informative scatter plots. However, the median errors (see the table within Fig.\ \ref{fig:onestepvsode}) are again close, and the outliers seem to be due to numerical error in the computation of $\Gamma^{-1}$ for quasi-singular $\Gamma$'s.

As a confirmation of the accuracy of the ODE method, we found the ``final'' match $M(\hAP(\tbf),\hGR(\ttr))$ [see Eq.\ \eqref{corr}] to be $> 0.9999$ for all sets of angles, and for all three waveform variants; final matches were considerably lower with the one-step formula (except for the hybrid waveforms, where they were $> 0.99$ for $93\%$ of the Monte Carlo population); and the ``initial'' matches $M(\hAP(\tbf),\hGR(\tbf))$ were always lower than $0.50$.

We conclude that the improved one-step formula yields \emph{at least} the correct order of magnitude for theoretical error, and usually does much better than that. In particular, it approximates well the theoretical errors for the intrinsic parameters and for the sky location, which is of special interest in view of the possibility of searching for electromagnetic counterparts to LISA MBHB sources. The one-step formula is attractive both because it is easy to implement and because it runs much faster than the ODE, by a factor equal approximately to the typical number of integration steps (in our work we found that $\sim 40$ steps were usually sufficient to achieve several digits of accuracy in $\ttr$). It is therefore appropriate to use the improved one-step formula in rapid Monte Carlo surveys over large parameter ranges, such as the survey discussed in the next section.

\subsection{Monte Carlo survey of theoretical errors}
\label{sec:survey}

We survey theoretical errors for eight representative MBHB systems defined by the component mass combinations
$(M_1 + M_2) = (10^5 + 10^5)M_\odot$,
$(10^4 + 10^6)M_\odot$,
$(10^5 + 10^6)M_\odot$,
$(10^6 + 10^6)M_\odot$,
$(10^4 + 10^7)M_\odot$,
$(10^5 + 10^7)M_\odot$,
$(10^6 + 10^7)M_\odot$, and
$(10^7 + 10^7)M_\odot$ \footnote{We had to exclude from our computation the analogous combination of lowest total mass, $M_1 + M_2 = (10^4 + 10^5)M_\odot$, which requires a higher Nyquist frequency (and therefore more points in the FFT representation) than MATLAB could support.}
and by $\beta = 0$. To do so, we compute improved--one-step theoretical errors for the straight, truncated, and hybrid signal models, for 600 random choices of sky-position and $\vec{L}$ angles uniformly distributed on the sky. In addition, we compute expected statistical errors from the Fisher-matrix formula [Eq.\ \eqref{bardx}], using 3.5PN waveforms truncated at $r = 6M$, and setting the $\mathrm{SNR} = 1000$. SNRs of this magnitude are indeed expected for the best LISA MBHB 
detections~\cite{Team:1998lr}, so they provide a reasonable benchmark to gauge the magnitude of theoretical errors.

\begin{figure}
\includegraphics[width=3.4in]{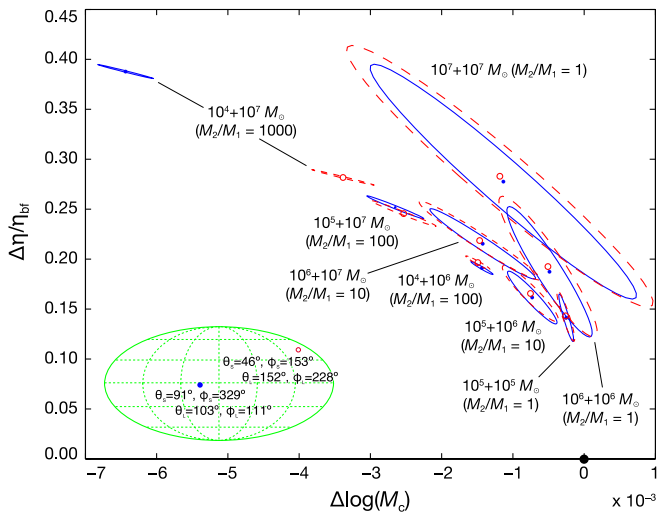}
\caption{Theoretical errors (dots) and $1\sigma$ statistical errors (ellipses) for several MBHB mass configurations, and for two representative sets of sky-position and $\vec{L}$ angles (near the ecliptic plane $\rightarrow$ solid dots and ellipses; closer to the pole $\rightarrow$ empty dots and dashed ellipses). For all systems $\beta = 0$ and $\mathrm{SNR} = 1000$. Strictly speaking, this plot displays \emph{minus} the theoretical error, since the dots mark the location of $\ttr$ relative to $\tbf$ (the thick dot at the origin). 
Statistical errors become smaller for more asymmetric mass configurations, and for smaller total masses.
The most asymmetric mass configuration, $(10^4+10^7)M_\odot$, displays the largest spread of $\ttr$. 
\label{fig:bubbles}}
\end{figure}

To visualize the relation between theoretical and statistical errors, consider Fig.\ \ref{fig:bubbles}: the thick dot at $\Delta \log M_c = \Delta \eta = 0$ marks the location $\tbf$ of the assumed best-fit parameters $\log M_c$ and $\eta$, while the small solid and empty dots show the location $\ttr$ of the true system parameters (computed for this figure with the ODE formula and the hybrid signal model), for two representative sky locations and $\vec{L}$ orientations, described at the bottom left of the figure. The ellipses around the dots enclose $1\sigma$ probability regions for $\ttr$ in the presence of noise-induced statistical error $\Delta\theta_n$ (computed from the Fisher matrix, and approximated as independent from $\Delta \theta_\mathrm{th}$ as in Sec.\ \ref{sec:linearized}). Recall that for a given $\theta_\mathrm{bf}$, $\Delta \theta_\mathrm{th}$ is deterministic (although it is unknown in practice, because we do not have access to the true $\hGR$), while $\Delta \theta_n$ is a random variable with distribution determined by the properties of the waveforms and of instrument noise.

In Tab.\ \ref{tab:bias} we show the median of the absolute theoretical errors over the populations of $\bar\theta_S$, $\bar\phi_S$, $\bar\theta_L$, and $\bar\phi_L$, for each representative mass combination, and for our three signal models. The first row in each group gives the median $1\sigma$ statistical errors.  In the last column, $A$ is defined as the combination $M_c^{5/6}/D$.  [Median absolute errors are more robust indicators of typical errors than the error variances, which can be corrupted (especially for $\Delta \bar\theta_L$, $\Delta \bar\phi_L$, $\Delta \Lambda_0$, and $\Delta \log A$) by the few outliers produced by the improved--one-step formula. The single-parameter $1\sigma$ statistical errors correspond in Fig.\ \ref{fig:bubbles} to half the projection of the ellipses on the horizontal and vertical axes.]
\begin{table*}
\caption{Median absolute theoretical errors in our survey of 600 sets of sky-position and $\vec{L}$ angles for eight representative mass combinations (with $\beta = 0$). The first line in each group gives the $1\sigma$ single-parameter statistical error at $\mathrm{SNR} = 1000$; the second to fourth lines give the median absolute theoretical errors for the straight, early-truncation, and hybrid signal models.\label{tab:bias}}
\begin{tabular}{r||l|l|l|l|l|l|l|l|l|l}
& \multicolumn{1}{c|}{$\Delta \log M_c$}
& \multicolumn{1}{c|}{$\Delta \eta / \eta$}
& \multicolumn{1}{c|}{$\Delta \beta$}
& \multicolumn{1}{c|}{$\Delta \bar\theta_S$}
& \multicolumn{1}{c|}{$\Delta \bar\phi_S$}
& \multicolumn{1}{c|}{$\Delta \bar\theta_L$}
& \multicolumn{1}{c|}{$\Delta \bar\phi_L$}
& \multicolumn{1}{c|}{$\Delta \Lambda_0$}
& \multicolumn{1}{c|}{$f_0 \Delta t_0$}
& \multicolumn{1}{c}{$\Delta \log A$} \\
\hline\hline
$(10^5+10^5)M_\odot$ stat. & $9.98\,10^{-6}$ & $2.29\,10^{-3}$ & $1.76\,10^{-2}$ & $1.96\,10^{-2}$ & $2.44\,10^{-2}$ & $3.04\,10^{-2}$ & $3.96\,10^{-2}$ & $9.55\,10^{-1}$ & $1.23\,10^{-1}$ & $2.45\,10^{-2}$ \\
          sys. & $4.47\,10^{-4}$ & $1.89\,10^{-1}$ & $1.18$ & $5.46\,10^{-2}$ & $8.94\,10^{-2}$ & $1.07\,10^{-1}$ & $1.55\,10^{-1}$ & $3.17$ & $1.58$ & $9.41\,10^{-2}$ \\
 (trunc.) sys. & $2.84\,10^{-4}$ & $1.40\,10^{-1}$ & $8.30\,10^{-1}$ & $2.51\,10^{-2}$ & $3.97\,10^{-2}$ & $4.64\,10^{-2}$ & $6.49\,10^{-2}$ & $7.84\,10^{-1}$ & $9.73\,10^{-1}$ & $3.88\,10^{-2}$ \\
 (hybrid) sys. & $5.86\,10^{-5}$ & $2.46\,10^{-2}$ & $1.53\,10^{-1}$ & $7.30\,10^{-3}$ & $1.22\,10^{-2}$ & $1.45\,10^{-2}$ & $2.11\,10^{-2}$ & $4.27\,10^{-1}$ & $2.02\,10^{-1}$ & $1.26\,10^{-2}$ \\
\hline
$(10^4+10^6)M_\odot$ stat. & $1.38\,10^{-5}$ & $5.82\,10^{-4}$ & $5.07\,10^{-3}$ & $1.49\,10^{-2}$ & $1.74\,10^{-2}$ & $1.87\,10^{-2}$ & $2.66\,10^{-2}$ & $1.57\,10^{-1}$ & $2.07\,10^{-2}$ & $1.67\,10^{-2}$ \\
          sys. & $3.59\,10^{-3}$ & $3.20\,10^{-1}$ & $2.04$ & $1.39\,10^{-1}$ & $1.80\,10^{-1}$ & $1.88\,10^{-1}$ & $2.76\,10^{-1}$ & $1.52$ & $2.89\,10^1$ & $1.53\,10^{-1}$ \\
 (trunc.) sys. & $1.68\,10^{-3}$ & $2.02\,10^{-1}$ & $1.13$ & $1.62\,10^{-2}$ & $2.09\,10^{-2}$ & $2.42\,10^{-2}$ & $3.32\,10^{-2}$ & $1.52\,10^{-1}$ & $1.79\,10^1$ & $2.23\,10^{-2}$ \\
 (hybrid) sys. & $2.00\,10^{-5}$ & $1.77\,10^{-3}$ & $1.13\,10^{-2}$ & $8.21\,10^{-4}$ & $1.08\,10^{-3}$ & $1.12\,10^{-3}$ & $1.60\,10^{-3}$ & $8.99\,10^{-3}$ & $1.59\,10^{-1}$ & $8.88\,10^{-4}$ \\
\hline
$(10^5+10^6)M_\odot$ stat. & $3.27\,10^{-5}$ & $2.53\,10^{-3}$ & $2.04\,10^{-2}$ & $2.53\,10^{-2}$ & $3.05\,10^{-2}$ & $3.66\,10^{-2}$ & $4.65\,10^{-2}$ & $2.64\,10^{-1}$ & $2.81\,10^{-2}$ & $2.89\,10^{-2}$ \\
          sys. & $1.90\,10^{-3}$ & $2.83\,10^{-1}$ & $1.74$ & $6.01\,10^{-2}$ & $7.21\,10^{-2}$ & $1.16\,10^{-1}$ & $1.23\,10^{-1}$ & $6.54\,10^{-1}$ & $3.80$ & $8.33\,10^{-2}$ \\
 (trunc.) sys. & $8.36\,10^{-4}$ & $1.68\,10^{-1}$ & $9.05\,10^{-1}$ & $8.42\,10^{-3}$ & $1.01\,10^{-2}$ & $1.29\,10^{-2}$ & $1.68\,10^{-2}$ & $6.27\,10^{-2}$ & $2.40$ & $1.01\,10^{-2}$ \\
 (hybrid) sys. & $8.70\,10^{-5}$ & $1.29\,10^{-2}$ & $7.92\,10^{-2}$ & $2.82\,10^{-3}$ & $3.46\,10^{-3}$ & $5.40\,10^{-3}$ & $5.85\,10^{-3}$ & $3.13\,10^{-2}$ & $1.71\,10^{-1}$ & $3.96\,10^{-3}$ \\
\hline
$(10^6+10^6)M_\odot$ stat. & $5.51\,10^{-5}$ & $6.25\,10^{-3}$ & $5.39\,10^{-2}$ & $1.94\,10^{-2}$ & $2.31\,10^{-2}$ & $2.89\,10^{-2}$ & $3.77\,10^{-2}$ & $1.22\,10^{-1}$ & $1.28\,10^{-2}$ & $2.50\,10^{-2}$ \\
          sys. & $1.40\,10^{-3}$ & $3.32\,10^{-1}$ & $2.32$ & $2.15\,10^{-2}$ & $1.89\,10^{-2}$ & $3.69\,10^{-2}$ & $4.39\,10^{-2}$ & $1.49\,10^{-1}$ & $1.46$ & $3.28\,10^{-2}$ \\
 (trunc.) sys. & $5.76\,10^{-4}$ & $1.89\,10^{-1}$ & $1.21$ & $3.18\,10^{-3}$ & $2.86\,10^{-3}$ & $4.89\,10^{-3}$ & $5.86\,10^{-3}$ & $1.38\,10^{-2}$ & $9.36\,10^{-1}$ & $3.93\,10^{-3}$ \\
 (hybrid) sys. & $1.61\,10^{-4}$ & $3.77\,10^{-2}$ & $2.64\,10^{-1}$ & $2.56\,10^{-3}$ & $2.26\,10^{-3}$ & $4.50\,10^{-3}$ & $5.26\,10^{-3}$ & $1.76\,10^{-2}$ & $1.64\,10^{-1}$ & $3.94\,10^{-3}$ \\
\hline
$(10^4+10^7)M_\odot$ stat. & $3.56\,10^{-5}$ & $6.66\,10^{-4}$ & $6.80\,10^{-3}$ & $1.46\,10^{-3}$ & $1.75\,10^{-3}$ & $2.40\,10^{-3}$ & $3.37\,10^{-3}$ & $4.81\,10^{-3}$ & $4.68\,10^{-2}$ & $2.47\,10^{-3}$ \\
          sys. & $1.44\,10^{-2}$ & $5.51\,10^{-1}$ & $4.33$ & $1.46\,10^{-2}$ & $1.63\,10^{-2}$ & $2.63\,10^{-2}$ & $3.82\,10^{-2}$ & $8.96\,10^{-2}$ & $2.62\,10^2$ & $1.81\,10^{-2}$ \\
 (trunc.) sys. & $7.22\,10^{-3}$ & $3.82\,10^{-1}$ & $2.70$ & $5.24\,10^{-3}$ & $8.27\,10^{-3}$ & $1.03\,10^{-2}$ & $1.64\,10^{-2}$ & $3.46\,10^{-2}$ & $1.54\,10^2$ & $4.64\,10^{-3}$ \\
 (hybrid) sys. & $6.65\,10^{-6}$ & $2.55\,10^{-4}$ & $2.00\,10^{-3}$ & $9.53\,10^{-6}$ & $1.12\,10^{-5}$ & $1.30\,10^{-5}$ & $1.68\,10^{-5}$ & $3.88\,10^{-5}$ & $1.21\,10^{-1}$ & $9.43\,10^{-6}$ \\
\hline
$(10^5+10^7)M_\odot$ stat. & $3.51\,10^{-5}$ & $1.04\,10^{-3}$ & $9.60\,10^{-3}$ & $2.29\,10^{-3}$ & $2.77\,10^{-3}$ & $3.93\,10^{-3}$ & $5.08\,10^{-3}$ & $6.89\,10^{-3}$ & $1.40\,10^{-2}$ & $3.57\,10^{-3}$ \\
          sys. & $6.54\,10^{-3}$ & $3.97\,10^{-1}$ & $2.78$ & $9.62\,10^{-3}$ & $8.22\,10^{-3}$ & $1.53\,10^{-2}$ & $2.08\,10^{-2}$ & $3.39\,10^{-2}$ & $2.83\,10^1$ & $9.23\,10^{-3}$ \\
 (trunc.) sys. & $3.36\,10^{-3}$ & $2.74\,10^{-1}$ & $1.73$ & $2.39\,10^{-3}$ & $2.44\,10^{-3}$ & $3.66\,10^{-3}$ & $5.01\,10^{-3}$ & $9.48\,10^{-3}$ & $1.70\,10^1$ & $2.43\,10^{-3}$ \\
 (hybrid) sys. & $3.37\,10^{-5}$ & $2.04\,10^{-3}$ & $1.43\,10^{-2}$ & $6.49\,10^{-5}$ & $5.12\,10^{-5}$ & $7.87\,10^{-5}$ & $1.12\,10^{-4}$ & $1.85\,10^{-4}$ & $1.45\,10^{-1}$ & $6.02\,10^{-5}$ \\
\hline
$(10^6+10^7)M_\odot$ stat. & $6.66\,10^{-5}$ & $3.23\,10^{-3}$ & $2.82\,10^{-2}$ & $4.69\,10^{-3}$ & $5.54\,10^{-3}$ & $6.96\,10^{-3}$ & $9.79\,10^{-3}$ & $1.11\,10^{-2}$ & $6.97\,10^{-3}$ & $6.46\,10^{-3}$ \\
          sys. & $3.45\,10^{-3}$ & $3.41\,10^{-1}$ & $2.29$ & $6.38\,10^{-3}$ & $5.06\,10^{-3}$ & $9.46\,10^{-3}$ & $1.32\,10^{-2}$ & $1.50\,10^{-2}$ & $3.75$ & $7.25\,10^{-3}$ \\
 (trunc.) sys. & $1.74\,10^{-3}$ & $2.28\,10^{-1}$ & $1.38$ & $1.44\,10^{-3}$ & $1.12\,10^{-3}$ & $2.29\,10^{-3}$ & $2.58\,10^{-3}$ & $3.46\,10^{-3}$ & $2.29$ & $1.63\,10^{-3}$ \\
 (hybrid) sys. & $1.49\,10^{-4}$ & $1.46\,10^{-2}$ & $9.85\,10^{-2}$ & $3.08\,10^{-4}$ & $2.42\,10^{-4}$ & $4.85\,10^{-4}$ & $5.77\,10^{-4}$ & $7.63\,10^{-4}$ & $1.59\,10^{-1}$ & $3.56\,10^{-4}$ \\
\hline
$(10^7+10^7)M_\odot$ stat. & $1.65\,10^{-4}$ & $1.08\,10^{-2}$ & $1.01\,10^{-1}$ & $6.81\,10^{-3}$ & $8.23\,10^{-3}$ & $1.18\,10^{-2}$ & $1.59\,10^{-2}$ & $1.57\,10^{-2}$ & $5.53\,10^{-3}$ & $9.89\,10^{-3}$ \\
          sys. & $3.13\,10^{-3}$ & $4.35\,10^{-1}$ & $3.31$ & $4.94\,10^{-3}$ & $3.28\,10^{-3}$ & $7.10\,10^{-3}$ & $1.07\,10^{-2}$ & $1.05\,10^{-2}$ & $1.43$ & $6.19\,10^{-3}$ \\
 (trunc.) sys. & $1.50\,10^{-3}$ & $2.87\,10^{-1}$ & $2.02$ & $1.05\,10^{-3}$ & $7.33\,10^{-4}$ & $1.83\,10^{-3}$ & $2.27\,10^{-3}$ & $2.47\,10^{-3}$ & $8.65\,10^{-1}$ & $1.39\,10^{-3}$ \\
 (hybrid) sys. & $3.29\,10^{-4}$ & $4.53\,10^{-2}$ & $3.45\,10^{-1}$ & $5.81\,10^{-4}$ & $3.87\,10^{-4}$ & $7.97\,10^{-4}$ & $1.27\,10^{-3}$ & $1.34\,10^{-3}$ & $1.47\,10^{-1}$ & $7.47\,10^{-4}$ \\
\hline\hline
\end{tabular}
\end{table*}

Several trends are apparent from Tab.\ \ref{tab:bias}. First, theoretical errors decrease (as expected) when we move from the straight signal models (truncated at $r = 6M$) to the early-truncation (at $r = 9M$) models.
The median improvement is only a factor $\sim 1.5\mbox{--}2.5$ for the intrinsic parameters $M_c$, $\eta$, and $\beta$, but $\sim 2\mbox{--}9$ for the angles $\bar\theta_S$, $\bar\phi_S$, $\bar\theta_L$, and $\bar\phi_L$. 
Theoretical errors are even smaller for our hybrid signal model, where $\hAP$ tends to $\hGR$ in the limit $\eta \rightarrow 0$. The improvement is most dramatic (factors of $10^2\mbox{--}10^3$) for mass combinations that actually have very small mass ratios, such as $(M_1 + M_2) = (10^4 + 10^6) M_{\odot}$ or $(10^5 + 10^7) M_{\odot}$.  But the improvement over the straight model is substantial (factors of $\sim 10$) even for the equal-mass cases.

While our estimated theoretical errors are generally small on an absolute scale, they are of comparable magnitude or larger than the $\mathrm{SNR} = 1000$ statistical errors, especially for the intrinsic parameters $M_c$, $\eta$, and $\beta$, and for $\Lambda_0$ and $t_0$ [with one exception for the hybrid model at the highest mass ratio, $(10^4+10^7)M_\odot$]. For the sky-position angles $\bar\theta_S$ and $\bar\phi_S$, 
theoretical and statistical errors are comparable when we use the straight and early-truncation signal models; however, $\Delta_\mathrm{th}\bar\theta_S$  and $\Delta_\mathrm{th}\bar\phi_S$ are always comfortably smaller than the statistical errors when we use the hybrid model.  
The results for $\bar\theta_L$, $\bar\phi_L$, and $\log A$ are quite similar to those for $\bar\theta_S$ and $\bar\phi_S$ (although, as mentioned above,  our one-step formula is less reliable for those parameters). 
 
\section{Summary and Future Work}  
\label{sec:conclusion}

The results of Sec.\ \ref{sec:survey} suggest that for hybrid PN waveforms, theoretical errors due to waveform inaccuracy will not be large enough to significantly degrade the available science from MBHBs. Nevertheless theoretical errors could well be the limiting factor in determining source parameters for the strongest sources.
This is especially true for the intrinsic parameters such as the masses and spins; theoretical errors appear relatively more benign (compared to statistical errors) for the sky position angles. [Also, to enable searches for electromagnetic counterparts to the inspiral GW signals, the sky position will have to be determined several days before the end of the inspiral, to allow for the intermittent transmission of LISA data to Earth, and to provide time to set up electromagnetic observations 
\cite{Kocsis:2007qy}. This will reduce the SNR available for detection (and therefore increase statistical error), but it will also reduce the theoretical error due to unknown high--PN-order terms, which are suppressed by fractional powers of the orbital frequency.]

The indications of this paper are tentative, in that the waveform models adopted in this paper are extremely simplified; several additional known effects, including amplitude modulations due to dynamically evolving spin--orbit and spin--spin couplings (see, e.g., Ref.\ \cite{PhysRevD.67.104025}), significant orbital eccentricity~\cite{Berti2006CQGra..23S.785B}, higher--PN-order GW harmonics \cite{2007PhRvD..75l4002A}, and post-inspiral parts of the waveform (i.e., merger and ringdown), modify the signal-space geometries of the template and true waveform manifolds in ways that are beneficial and ways that are damaging for parameter estimation; the net effect needs to be computed.
Improved surveys of theoretical error that include the waveform features listed above can be performed with the mathematical tools described in Sec.\ \ref{sec:parameterestimation} of this paper. This
will be very important in establishing whether further work in PN and NR calculations is needed for LISA to reach its full scientific potential (ultimately limited by the source-confusion and instrument noise profiles, and not by the development of GR theory). 

The fact that the match $M(\hAP(\tbf),\hGR(\ttr))$ turns out to be $> 0.9999$ for all three approximate waveforms is a rather striking result.  If this continues to hold for more realistic waveform models, it will mean that it may be quite hard (for typical SNRs) to judge, from goodness of fit to the data, how accurate one's theoretical template waveforms really are.
 

There are other applications of those methods. For instance, almost the same problem studied for LISA in this paper arises in the context of ground-based GW astronomy, 
in estimating parameters from the inspirals and mergers of binaries of neutron stars and/or stellar-mass BHs. Also, there are many cases in GW observations where theoretical approximations are made to simplify calculations (again effectively modifying the ``search'' template family with the respect to the true measured signals), and the tools of this paper could be used to investigate the impact of these approximations on the extraction of source information from observations. One such approximation, recently critiqued by Grishchuk \cite{2004CQGra..21.4041B}, is the long-wavelength formula for the LIGO and VIRGO GW response functions, which introduces distortions at frequencies of the order of the inverse light-travel time across the interferometer arms.  Another application is to LISA searches for extreme-mass-ratio inspirals (EMRIs), where $\eta$ is typically $\sim 10^{-5}$.  A great deal of effort has been expended, and impressive progress made, in calculating EMRI waveforms to first-order in an expansion in $\eta$ \cite{2004LRR.....7....6P}. How important is it for theorists to calculate EMRI waveforms through order $\eta^2$~\cite{2006PhRvD..74h4018R}?  We plan to address that question in a future paper.


\acknowledgments

We would like to thank A.\ Buonanno, Y.\ Chen, E.\ Berti, and B.\ Kocsis for helpful discussions.  We thank A.\ Vecchio and S.\ Hughes for graciously sharing 
with us their MBHB parameter-estimation codes (although we ended up not using them for this paper). CC's and MV's work was carried out at the Jet Propulsion Laboratory, California Institute of Technology, under contract to the National Aeronautics and Space Administration. MV is grateful for support from the Human Resources Development Fund program at JPL. 

\newcommand{\mnras}{Mon. Not. Roy. Astron. Soc.}
\newcommand{\aap}{A\&A}

\end{document}